\documentclass{elsart}

\IfFileExists{srcltx.sty}{\usepackage[active]{srcltx}} 

\usepackage{amsmath,amssymb,epsfig}

\journal{Physics Reports}

\begin{document}

%%%%%%%%%%%%%%%%%%%%%%%%%%%%%%%%%%%%%%%%
% Own definitions                      %
%%%%%%%%%%%%%%%%%%%%%%%%%%%%%%%%%%%%%%%%

% Units

\newcommand{\eV}{\text{eV}}
\newcommand{\keV}{\text{keV}}
\newcommand{\MeV}{\text{MeV}}
\newcommand{\GeV}{\text{GeV}}
\newcommand{\arcmin}{\text{arcmin}}
\newcommand{\Mpc}{\text{Mpc}}
\newcommand{\Hunit}{$\text{km\,s}^{-1}\,\text{Mpc}^{-1}$}
\newcommand{\muK}{$\mu\text{K}$}
%%%%%%%%%%%%%%%
\newcommand{\beq}{\begin{equation}}
\newcommand{\eeq}{\end{equation}}
\newcommand{\bea}{\begin{eqnarray}}
\newcommand{\eea}{\end{eqnarray}}
\newcommand{\dd}{\partial}
\newcommand{\greeksym}[1]{{\usefont{U}{psy}{m}{n}#1}}
\newcommand{\umu}{\mbox{\greeksym{m}}}
\newcommand{\udelta}{\mbox{\greeksym{d}}}
\newcommand{\uDelta}{\mbox{\greeksym{D}}}
\newcommand{\uPi}{\mbox{\greeksym{P}}}
\newcommand{\F}{\Phi}
\newcommand{\f}{\phi}
\newcommand{\vf}{\varphi}
\newcommand{\Q}{\tilde{Q}_{_L}}
\newcommand{\q}{\tilde{q}_{_R}}
\newcommand{\Lp}{\tilde{L}_{_L}}
\newcommand{\lp}{\tilde{l}_{_R}}
\newcommand{\nL}{$\, \backslash \! \! \! \! L \ $}
%%%%%%%%%%%%%%%%%%%%%%%%%%%%%%%%%
\newcommand{\mh}{m_{_{\rm H}}}
\newcommand{\ms}{m_{_{\rm S}}}
\newcommand{\lh}{\lambda_{_{\rm H}}}
\newcommand{\ls}{\lambda_{_{\rm S}}}
\newcommand{\lm}{\lambda_{_{\rm HS}}}
%
% Derivatives

\newcommand{\D}{\text{D}}
\newcommand{\ud}{\text{d}}
\newcommand{\curl}{\,\text{curl}\,}
\newcommand{\slashed}[1]{{#1}\hspace{-2mm}/}
% Inequalities

\newcommand{\alt}{\lesssim}
\newcommand{\agt}{\gtrsim}

% Miscellaneous cosmology

\newcommand{\Omtot}{\Omega_{\mathrm{tot}}}
\newcommand{\Omb}{\Omega_{\mathrm{b}}}
\newcommand{\Omc}{\Omega_{\mathrm{c}}}
\newcommand{\Omm}{\Omega_{\mathrm{m}}}
\newcommand{\omb}{\omega_{\mathrm{b}}}
\newcommand{\omc}{\omega_{\mathrm{c}}}
\newcommand{\omm}{\omega_{\mathrm{m}}}
\newcommand{\omnu}{\omega_{\nu}}
\newcommand{\Omnu}{\Omega_{\nu}}
\newcommand{\Oml}{\Omega_\Lambda}
\newcommand{\OmK}{\Omega_K}

% Miscellaneous HD & MHD

%\newcommand{\Cs}{c_{\rm s}}
%\newcommand{\Cs2}{c_{\rm s}^2}
%\newcommand{\Ca}{c_{\rm a}}
%\newcommand{\Ca2}{c_{\rm a}^2}

% Cross sections

\newcommand{\sigt}{\sigma_{\mbox{\scriptsize T}}}

% Journals
% 
\newcommand{\annphys}{\rm Ann.~Phys.~}
\newcommand{\araa}{\rm Ann.~Rev.~Astron.~\&~Astrophys.~}
\newcommand{\aap}{\rm Astron.~\&~Astrophys.~}
\newcommand{\apj}{\rm Astrophys.~J.~}
\newcommand{\apjl}{\rm Astrophys.~J.~Lett.~}
\newcommand{\apjs}{\rm Astrophys.~J.~Supp.~}
\newcommand{\apss}{\rm Astrophys.~Space~Sci.~}
\newcommand{\cqg}{\rm Class.~Quant.~Grav.~}
\newcommand{\grg}{\rm Gen.~Rel.~Grav.~}
\newcommand{\ijmpd}{\rm Int.~J.~Mod.~Phys.~D~}
\newcommand{\jcap}{\rm JCAP~}
\newcommand{\jetpl}{\rm J.~Exp.~Theor.~Phys.~Lett.~}
\newcommand{\jmp}{\rm J.~Math.~Phys.~}
\newcommand{\mnras}{\rm Mon.~Not.~R.~Astron.~Soc.~}
\newcommand{\nat}{\rm Nature~}
\newcommand{\prd}{\rm Phys.~Rev.~D~}
\newcommand{\prl}{\rm Phys.~Rev.~Lett.~}
\newcommand{\plb}{\rm Phys.~Lett.~B~}
\newcommand{\physrep}{\rm Phys.~Rep.~}
\newcommand{\progthp}{\rm Prog.~Theor.~Phys.~}
\newcommand{\rmp}{\rm Rev.~Mod.~Phys.~}
\newcommand{\jgr}{\rm J.~Geophys.~Res.}

\let\Oldsection\section
\renewcommand{\section}[1]{\Oldsection{\bf #1}}

\newcommand{\be} {\begin{equation}}
\newcommand{\ee} {\end{equation}}
% \newcommand{\bea} {\begin{eqnarray}}
% \newcommand{\eea} {\end{eqnarray}}

%%%%%%%%%%%%%%%%%%%%%%%%%%%%%%%%%%%%%%%%
% Front matter                         %
%%%%%%%%%%%%%%%%%%%%%%%%%%%%%%%%%%%%%%%%

\begin{frontmatter}

\date{}
\title{{\small \mbox{} \hfill \rm \small UCLA/09/TEP/55}\\[1.5ex]
Sterile neutrinos: \\ the dark side of the light fermions}


\author{Alexander Kusenko}
\address{Department of Physics and Astronomy, University of California, Los
Angeles, CA 90095-1547, USA 
}
\address{Institute for the Physics and Mathematics of the Universe,
University of Tokyo, Kashiwa, Chiba 277-8568, Japan}

\begin{abstract}
The discovery of neutrino masses suggests the likely existence of gauge singlet fermions that participate in the neutrino mass generation via the seesaw mechanism.  The masses of the corresponding degrees of freedom can range from well below the electroweak scale to the Planck scale.   If some of the singlet fermions are light, the sterile neutrinos appear in the low-energy effective theory.  They can play an important role in astrophysics and cosmology.  In particular, sterile neutrinos with masses of several keV can account for cosmological dark matter, which can be relatively warm or cold, depending on the production mechanism.  The same particles can explain the observed  velocities of pulsars because of  the anisotropy in their emission from a cooling neutron star born in a supernova explosion.  Decays of the relic sterile neutrinos can produce a flux of X-rays that can affect the formation of the first stars.  Existing and future X-ray telescopes can be used to search for the relic sterile neutrinos.  
\end{abstract}

% \begin{keyword}
% neutrino masses, sterile neutrinos, dark matter, supernova physics, pulsars 
% \end{keyword}
\end{frontmatter}
\newpage
\tableofcontents

\newpage

%%%%%%%%%%%%%%%%%%%%%%%%%%%%%%%%%%%%%%%%
% Main text                            %
%%%%%%%%%%%%%%%%%%%%%%%%%%%%%%%%%%%%%%%%

\section{Introduction}

A  discovery in particle physics usually amounts to either a measurement of some
parameter related to a known particle, or a detection of some new degrees of
freedom,  new particles.  One could argue that the discovery of the neutrino
mass~\cite{Strumia:2006db} is likely to be both.  Not only is it a measurement of the non-zero mass,
but it also implies the likely existence of some additional, SU(3)$\times$SU(2)$\times$U(1) singlet fermions,
``right-handed'' neutrinos.  The corresponding particles can be made very heavy even for small
masses of the active neutrinos (the seesaw
mechanism~\cite{Minkowski:1977sc,Glashow:1979nm,Yanagida:1979as,GellMann:1980vs,Mohapatra:1979ia}),
but they can also be light, in which case they are called sterile neutrinos.  The above conclusion
is not a mathematical theorem, because  the Majorana masses of the active
neutrinos could emerge as SU(2) triplet higher-dimension operators generated at a high scale
due to string theory or other dynamics not involving the singlet fermions.  However, the ease with
which the seesaw mechanism can generate these masses makes the seesaw mechanism very appealing in its simplicity.  This mechanism can explain the smallness of neutrino masses even if the Yukawa couplings are
large.  However, there is no fundamental theory of the Yukawa couplings, and, as discussed below,
the naturalness arguments can be made in favor of either the large or the small Yukawa couplings. 
In this review we will consider the astrophysical and cosmological ramifications of the seesaw
models in which the Yukawa couplings and the right-handed Majorana mass terms are allowed 
to be relatively small.  In this case, the new degrees of freedom, introduced with the gauge singlet
fermions, appear in the low-energy effective theory, below the electroweak scale and can have a
number of important observable consequences, from dark matter to  pulsar kicks and other
astrophysical phenomena. 

As usual, we will call the neutrino weak eigenstates lefthanded  if they
transform as doublets under the SU(2) gauge group of the Standard Model.  The gauge singlet
fermions in the weak eigenstates basis are called singlet or righthanded neutrinos.  The mass eigenstates corresponding to masses below the electroweak scale will generally have small mixing angles and split into two classes: (i) those that couple to $Z$ boson with the coupling approximately equal to the
electron's coupling to $Z$, and (ii) those that will have a much smaller coupling to $Z$
(suppressed by the mixing parameter smaller than $10^{-2}$).  The former are called  
{\em active neutrinos}, and the latter are called {\em sterile neutrinos}\footnote{The name {\em
sterile neutrino} was coined by Bruno~Pontecorvo in a seminal
paper~\cite{Pontecorvo:1967fh}, in which he also considered vacuum neutrino
oscillations in the laboratory and in astrophysics, the lepton number
violation, the neutrinoless double beta decay, some rare processes, such as
$\mu \rightarrow e \gamma$, and several other questions that have dominated the
neutrino physics for the next four decades.}. 

\section{Neutrino masses and the emergence of sterile neutrinos}
\label{sec_masses}

The Standard Model was originally formulated with 
massless neutrinos $\nu_\alpha$ transforming as components of the
electroweak SU(2) doublets $L_\alpha$ ($\alpha =e,\mu,\tau$).  To accommodate the
neutrino masses, one can add several electroweak
singlets $ N_{a}$ ($a=1,...,n$) to build a
seesaw
Lagrangian~\cite{Minkowski:1977sc,Glashow:1979nm,Yanagida:1979as,GellMann:1980vs,Mohapatra:1979ia}:
\beq  {\mathcal L} = {\mathcal L_{\rm SM}} + i \bar
N_a \slashed{\partial} N_a - y_{\alpha a} H^{\dag} \,  \bar L_\alpha
N_a - \frac{M_a}{2} \; \bar N_a^c N_a + h.c.
\label{lagrangianM}
\eeq
Here ${\mathcal L_{\rm SM}} $ is the Standard Model Lagrangian (with only the left-handed neutrinos and without the neutrino masses).  We will assume that SU(3)-triplet Higgs bosons~\cite{Schechter:1980gr} are not involved, and all the neutrino masses arise from the Lagrangian~(\ref{lagrangianM}).  

\subsection{The seesaw mechanism}

The neutrino mass eigenstates $\nu^{\rm (m)}_i$ ($i=1,...,n+3$) are linear
combinations of the weak eigenstates $\{\nu_\alpha, N_a \}$.  They are obtained
by diagonalizing the $(n+3)\times (n+3)$ mass matrix:
\beq 
{\mathcal M^{(n+3)}} = \left( \begin{array}{cc}
          0 & y_{\alpha a} \langle H \rangle \\
y_{a \alpha} \langle H \rangle & {\rm diag}\{M_1,...,M_n\}
         \end{array}
\right).
\label{massmatrix_full}
\eeq
As long as all $y_{a \alpha} \langle H \rangle \sim y \langle H \rangle \ll M_a\sim M$, the eigenvalues of
this matrix split into two groups:  the lighter states with masses 
\begin{equation}
m(\nu^{\rm (m)}_{1,2,3} )  \sim  \frac{y ^2 \langle H \rangle ^2}{M}  
\end{equation}
and the heavier eigenstates with masses of the order of $M$: 
\begin{equation}
m(\nu^{\rm (m)}_{a} )  \sim  M \ \ \ (a>3).
\end{equation}
We call the former {\em active 
neutrinos} and the latter {\em sterile neutrinos}.  Generically, the mixing angles in this
case are of the order of 
\begin{equation}
\theta_{a\alpha}^2 \sim \frac{y_{a \alpha} ^2 \langle H
\rangle ^2}{M^2},
\end{equation}
but some additional symmetries or accidental cancellations can make them different from these generic values.

One can consider a broad range of values for the number $n$ of sterile neutrinos.  Unlike the
other fermions, the singlets are not subject to any constraint based on the anomaly
cancellation because these fermions do not couple to the gauge fields.  To explain
the neutrino masses inferred from the atmospheric and solar neutrino experiments, $n=2$ singlets
are sufficient~\cite{Frampton:2002qc}, but a greater number is required if the
Lagrangian (\ref{lagrangianM}) is to explain the r-process
nucleosynthesis~\cite{McLaughlin:1999pd}, the pulsar
kicks~\cite{Kusenko:1997sp,Kusenko:1998bk,Fuller:2003gy,Barkovich:2004jp,Kusenko:2004mm,Loveridge:2003fy,Kusenko:2006rh,Kusenko:2008gh} and the strength of the supernova
explosion~\cite{Fryer:2005sz,Hidaka:2006sg}, as well as dark  matter~\cite{Dodelson:1993je,Shi:1998km,Abazajian:2001nj,Abazajian:2001vt,Abazajian:2002yz,Dolgov:2000ew,Asaka:2005an,Kishimoto:2006zk,Asaka:2006ek,Asaka:2006rw}.   A model often referred to as  $\nu$MSM, for Minimal Standard Model (MSM) with neutrino ($\nu$) masses is the above model with $n=3$ sterile neutrinos, all of which have masses below the electroweak scale: one has a mass of the order of a few  keV, while the two remaining sterile neutrino are assumed to be closely degenerate at about 1-10~GeV scale~\cite{Asaka:2005an}.  This model is singled out for the minimal particle content consistent with baryogenesis~\cite{Akhmedov:1998qx,Asaka:2005pn} and having a dark-matter candidate.  However, as discussed below, the need for a cosmological  mechanism capable of producing colder dark matter than that generated by neutrino oscillations may require one to go beyond the minimal model, and introduce some additional physics at the electroweak scale~\cite{Kusenko:2006rh,Petraki:2007gq}.

The scale of the right-handed Majorana masses $M_{a}$ is unknown; it can be
much greater than the electroweak
scale~\cite{Minkowski:1977sc,Glashow:1979nm,Yanagida:1979as,GellMann:1980vs,Mohapatra:1979ia}, or it
may be as low as a few eV~\cite{deGouvea:2005er}.  Theoretical arguments have been put forth for various ranges of these Majorana masses.

\subsection{What is more natural?} 

The seesaw mechanism~\cite{Minkowski:1977sc,Glashow:1979nm,Yanagida:1979as,GellMann:1980vs,Mohapatra:1979ia} can explain the smallness of the neutrino
masses in the presence of the Yukawa couplings of order one if the
Majorana masses $M_a$ are much larger than the electroweak scale. Indeed, in
this case the masses of the lightest neutrinos are suppressed by the ratios $
\langle H \rangle/M_a$.  

However, the origin of the Yukawa couplings remains unknown, and there is no
experimental evidence to suggest that these couplings must be of order 1. In
fact, the Yukawa couplings of the charged leptons are much smaller than 1. For
example, the Yukawa coupling of the electron is as small as $10^{-6}$.  
One can ask whether some theoretical models are more likely to produce the
numbers of order one or much smaller than one.  The two possibilities are, in
fact, realized in two types of theoretical models.  If the Yukawa couplings
arise as some topological intersection numbers in string theory, they are
generally expected to be of order one~\cite{Candelas:1987rx}, although very
small couplings are also possible~\cite{EytonWilliams:2005bg}. If the Yukawa
couplings arise from the overlap of the wavefunctions of fermions located on
different branes or in the bulk in extra dimensions, they can be exponentially suppressed and
are expected to be very small~\cite{ArkaniHamed:1998vp,Mirabelli:1999ks,Hebecker:2002xw}.

In the absence of the fundamental theory, one may hope to gain some insight 
about the size of the Yukawa couplings using 't~Hooft's naturalness
criterion~\cite{'tHooft:1979bh}, which states that a number can be
naturally small if setting it to zero increases the symmetry of the Lagrangian.
 A small breaking of the symmetry is then associated with the small non-zero
value of the parameter.  This naturalness criterion has been applied to a
variety of theories; it is, for example, one of the main arguments in favor of
supersymmetry. (Setting the Higgs mass to zero does not increase the symmetry
of the Standard Model.  Supersymmetry relates the Higgs mass to the Higgsino
mass, which is protected by the chiral symmetry.  Therefore, the light Higgs
boson, which is not natural in the Standard Model, becomes natural in theories
with softly broken supersymmetry.) In view of 't~Hooft's criterion, the
\textit{small} Majorana mass is natural because setting $M_a$ to zero increases
the symmetry of the Lagrangian
(\ref{lagrangianM})~\cite{Fujikawa:2004jy,deGouvea:2005er}.  

One can also ask whether cosmology can provide any clues as to whether the mass
scale of sterile neutrinos should be above or below the electroweak scale.  It
is desirable to have a theory that could generate the matter--antimatter
asymmetry of the universe. In both limits, of large and small $M_a$, one can have
a successful leptogenesis: in the case of the high-scale seesaw, the baryon
asymmetry can be generated from the out-of-equilibrium decays of heavy
neutrinos~\cite{Fukugita:1986hr}, while in the case of the low-energy seesaw,
the matter-antimatter asymmetry can be produced by  
the neutrino oscillations~\cite{Akhmedov:1998qx,Asaka:2005pn}.  
Big-bang nucleosynthesis (BBN) can provide a constraint on the number of light
relativistic species in equilibrium~\cite{Barbieri:1989ti,Kainulainen:1990ds,Kirilova:1997sv,Shi:1998xs,McLaughlin:1999pd,Caldwell:1999zk}, but the sterile neutrinos
with the small mixing angles may never be in equilibrium in the early universe,
even at the highest temperatures~\cite{Dodelson:1993je}.  Indeed, the effective mixing angles 
of neutrinos at high temperature are suppressed due to the interactions with
plasma~\cite{Stodolsky:1986dx,Barbieri:1989ti,Kainulainen:1990ds,Barbieri:1990vx}, and, therefore, the sterile neutrinos may never thermalize.  High-precision measurements of the primordial abundances may probe
the existence of sterile neutrinos and the lepton asymmetry of the universe in
the future~\cite{Smith:2006uw}.  

While many seesaw models assume that the sterile neutrinos have very large
masses, which makes them unobservable (except, via leptogenesis), it is, of course, worthwhile to consider light sterile neutrinos in view of the above arguments, and also because they can
explain several experimental results, such as dark matter~\cite{Dodelson:1993je}, the 
velocities of pulsars~\cite{Kusenko:1997sp,Fuller:2003gy,Kusenko:2004mm},
etc.~\cite{Biermann:2006bu,Stasielak:2006br,Stasielak:2007ex,Stasielak:2007vs}. 
Sterile neutrinos with mass $\sim 10$~keV and a small mixing angle are predicted
by some models of neutrino masses~\cite{Ma:2009gu,Ma:2009kh,Cogollo:2009yi}. In
particular, electroweak-singlet fermions with masses below the electroweak
scale are predicted in some models of dynamical symmetry
breaking~\cite{Appelquist:2002me,Appelquist:2003uu,Appelquist:2003hn}.

\subsection{Split scales}

The Majorana masses $M_a$ may, of course, differ in scale.  For example, all but one of them may be at some high scale, while one Majorana mass may be below the electroweak scale.   In this case it is useful to integrate out the high-scale states from the full mass matrix in eq.~(\ref{massmatrix_full}).  Then in the low-energy effective theory one recovers the mass matrix of three active neutrinos and one sterile neutrino (see, e.g., Ref.~\cite{Smirnov:2006bu} and references therein): 
\begin{equation}
{\mathcal M}^{(4\times 4)}=
\left(\begin{array}{cc}
{\mathcal M}^{(3\times 3)} & \mathcal{M}^{(1\times 3)}_{\alpha 1}\\
\mathcal{M}^{(1\times 3)T}_{\alpha 1} & M_{1}
\end{array}
\right),
\label{total-m}
\end{equation}
where
\begin{equation}
{\mathcal M}^{(3\times 3)} =\left(\begin{array}{ccc} 
m_{ee} & m_{e\mu} & m_{e\tau}\\
m_{\mu e} & m_{\mu \mu} & m_{\mu \tau}\\
m_{\tau e } & m_{\tau \mu } & m_{\tau \tau}
\end{array}
\right)
\label{active_m}
\end{equation}
is the mass matrix of active neutrinos generated  by the see-saw mechanism after the high-scale states have been integrated out, and $ \mathcal{M}^{(1\times 3)}_{\alpha 1}=\{m_{\alpha 1} \}$, $(\alpha=e,\mu,\tau)$.  The patterns of masses and mixing angles among the active neutrinos can be affected by the terms induced by the presence of sterile neutrinos in the mass matrix of eq.~(\ref{total-m}).  Smirnov and Funchal~\cite{Smirnov:2006bu} investigated the possible effects of such induced mass terms as a function of sterile neutrino mass and the active-sterile mixing.  In some ranges of parameters such induced contribution could be responsible the observed  peculiar features of the effective mass matrix of active neutrinos~\cite{Smirnov:2006bu}.

The cosmological bounds on the sum of the active neutrino masses~\cite{Ichiki:2008ye} imply that the sum of the eigenvalues of mass matrix (\ref{total-m}) corresponding to the three (mostly) active eigenstates have to be  less than 0.54~eV.  These eigenvalues, in the limit of small active-sterile mixing, are close to the eigenvalues of the matrix (\ref{active_m}).  Unless some cancellations occur (which is possible, especially, in the presence of some new symmetries),  one can  consider $m_a \leq 1$~eV.   One can also assume that $M_1\gg m_{a1}$ and $M_1\gg m_{\alpha \beta}$.  Then all the active-sterile mixing angles are small, and the sterile neutrino has mass close to $M_1$. In most of our discussion, we will concentrate on this one sterile neutrino with mass 
\begin{equation}
 m_s\approx M_1.
\end{equation}
At the same time, one should keep in mind that some additional light sterile neutrinos may exist; they can play an important role in baryogenesis~\cite{Akhmedov:1998qx,Asaka:2005pn} and can alter the production rates of the dark-matter sterile neutrino in some cases~\cite{Laine:2008pg}.

\subsection{Can the Majorana masses originate at the electroweak-scale?}

Although the Majorana mass at the keV scale could be a fundamental parameter, it can also arise, as the other fermion masses, from the Higgs mechanism~\cite{Chikashige:1980ht}.  The interaction of the form 
\begin{equation} 
{\mathcal L} \supset \frac{h_a}{2} \, S \,
\bar {N}_{a}^c N_{a} + h.c. \,, 
\label{adding_S}
\end{equation}
where $S$ can be either real or complex, can generate the Majorana mass after the boson $S$ acquires a VEV.  If $S$ is complex, a U(1) symmetry (lepton number) is spontaneously broken by the VEV of $S$, and the Nambu-Goldstone boson, called Majoron, appears in the low-energy effective Lagrangian~\cite{Chikashige:1980ht}.  This theory does not contain a sterile neutrino that is stable enough to be dark matter because the sterile neutrino can decay into the Majoron and an active neutrino.   If the SU(2)-singlet $S$ is real, then the coupling in eq.~(\ref{adding_S}) breaks the lepton number explicitly (just like the mass term in eq.~(\ref{lagrangianM})), and no light scalar appears in the theory.  Hence, sterile neutrino does not have a fast decay channel involving a Majoron, and this sterile neutrino is stable enough to be dark matter.  The VEV and mass of this scalar will be discussed below.  A light, sub-GeV scalar could play the role of the inflaton~\cite{Shaposhnikov:2006xi}.  However, if one requires that the mass and the VEV be of the same order of magnitude, and that the dark matter density be acceptable, then one finds that the singlet ends up right at the electroweak scale~\cite{Kusenko:2006rh,Petraki:2007gq}, where it can have important implications for LHC phenomenology, the electroweak phase transition, and  baryogenesis~\cite{Kainulainen:1990ds,Enqvist:1992va,McDonald:1993ey,Vilja:1993uw,Datta:1995qw,Ham:2004cf,Ahriche:2007jp,Barger:2006sk,Barger:2007im,Profumo:2007wc}.

One can consider the following scalar potential:
\begin{equation}
V(H,S) =   m_{1}^2 |H|^2 + m_{2}^2 S^2+ \lambda_3 S^3 +  
\lambda_{_{HS}} |H|^2 S^2+ \lambda_{_S}  S^4   + \lambda_{_H}
|H|^4 ,
\label{potential}
\end{equation}
which allows both $H$ and $S$ to have non-zero VEV.  After the electroweak symmetry breaking, the Higgs doublet and the singlet, each develop a VEV, 
\beq
\langle H\rangle= v_0=247 \ {\rm GeV}, \ \langle S\rangle= v_1,
\eeq
and the singlet fermions acquire the Majorana masses 
\beq
 M_a = h_a v_1.
\eeq
As discussed below, this model
is suitable for generating dark matter in the form of sterile neutrinos, and it has two very
appealing features.  First, dark matter generated at the electroweak scale is not as warm as that
produced in oscillations, hence it probably is in better agreement with the data.  Second, the
dark matter abundance, $ \Omega_{\rm DM}\sim 0.2$ is a natural consequence of setting the scale of
the Higgs singlet, $\langle S\rangle= 10^2$~GeV, that is at the electroweak scale.

\section{Experimental status}

\begin{figure}[ht]
\centerline{\epsfxsize=5in\epsfbox{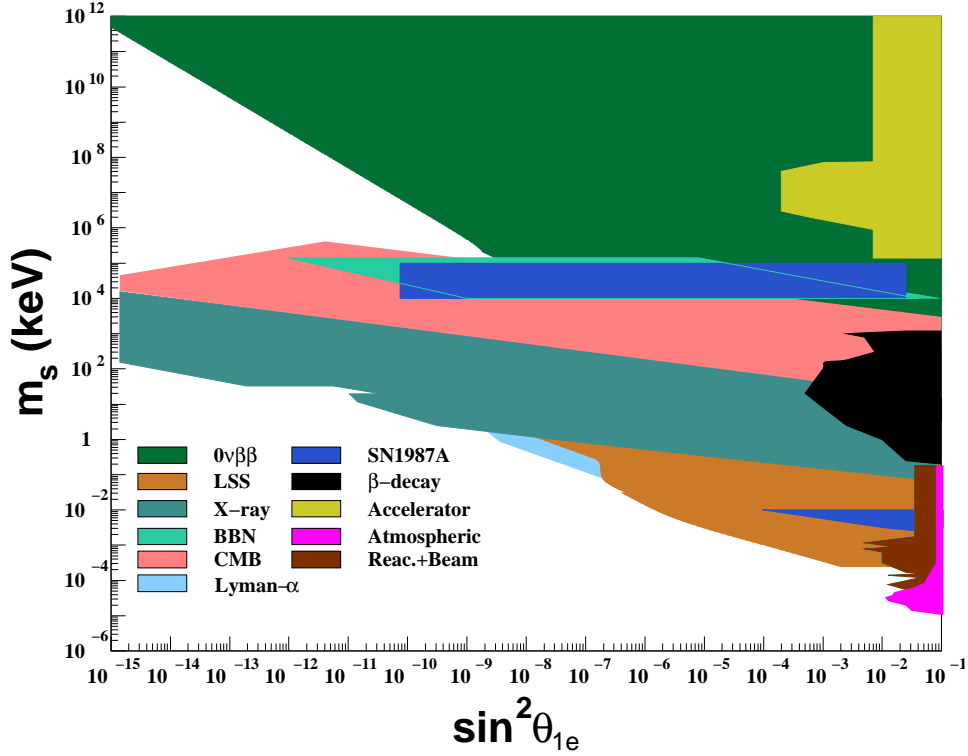}}   
\caption{ Experimental and observational limits on sterile neutrinos that have a
non-zero mixing with the electron neutrino only.  This figure is based on the
limits from Refs.~\cite{Kusenko:2004qc,Smirnov:2006bu,Atre:2009rg}.   The X-ray
limits and Lyman-$\alpha$ limits shown here are based on the abundances of relic
neutrinos produced by neutrino oscillations for zero lepton asymmetry; see
discussion in the text and in 
Refs.~\cite{Kusenko:2006rh,Palazzo:2007gz,Boyarsky:2008xj,Boyarsky:2008mt}
}
\label{figure:limits_e}
\end{figure}

\begin{figure}[ht]
\centerline{\epsfxsize=5in\epsfbox{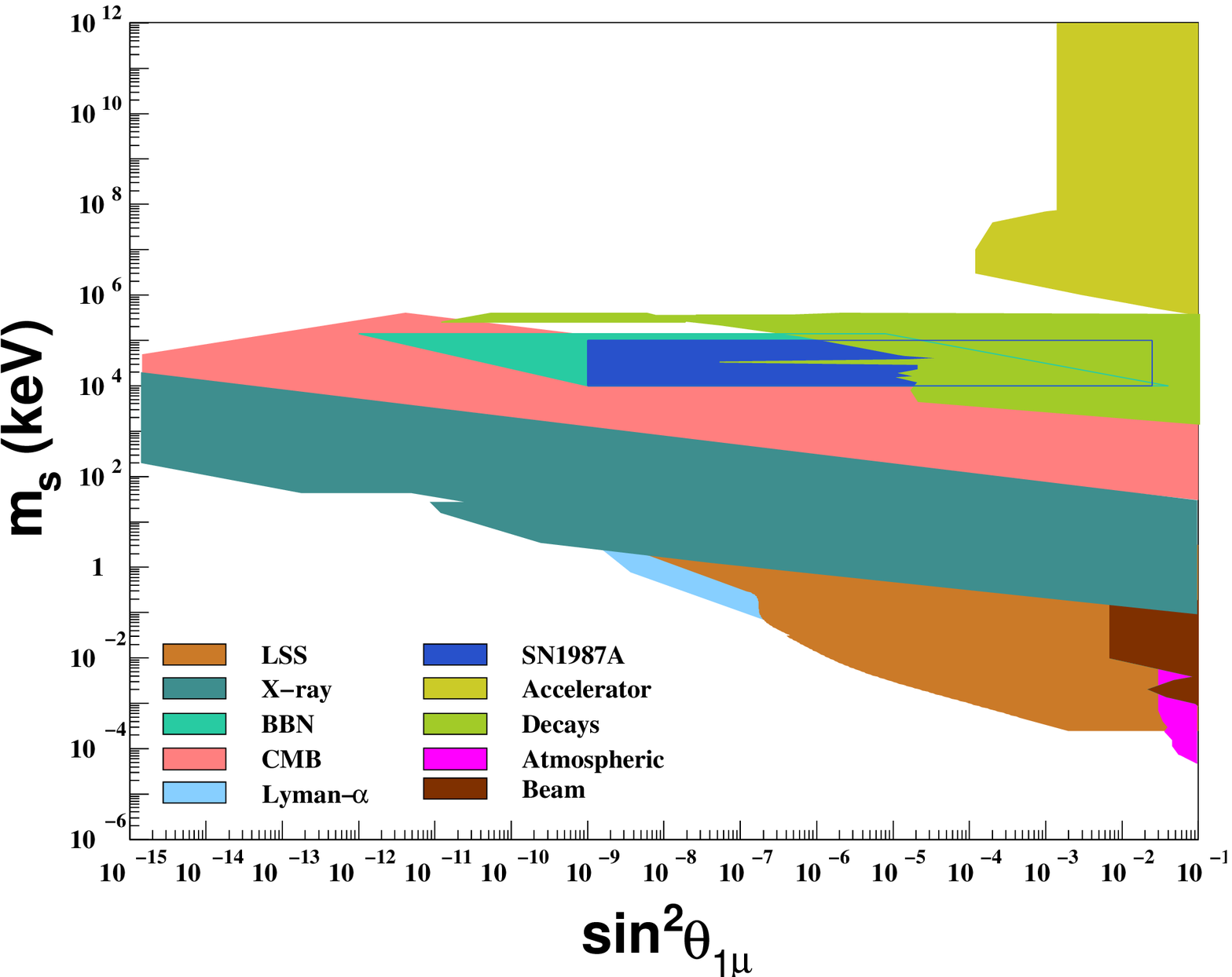}}   
\caption{ Same as in Fig.~\ref{figure:limits_e}, for the sterile neutrino mixing with $\nu_\mu$ only. 
}
\label{figure:limits_mu}
\end{figure}

\begin{figure}[ht]
\centerline{\epsfxsize=5in\epsfbox{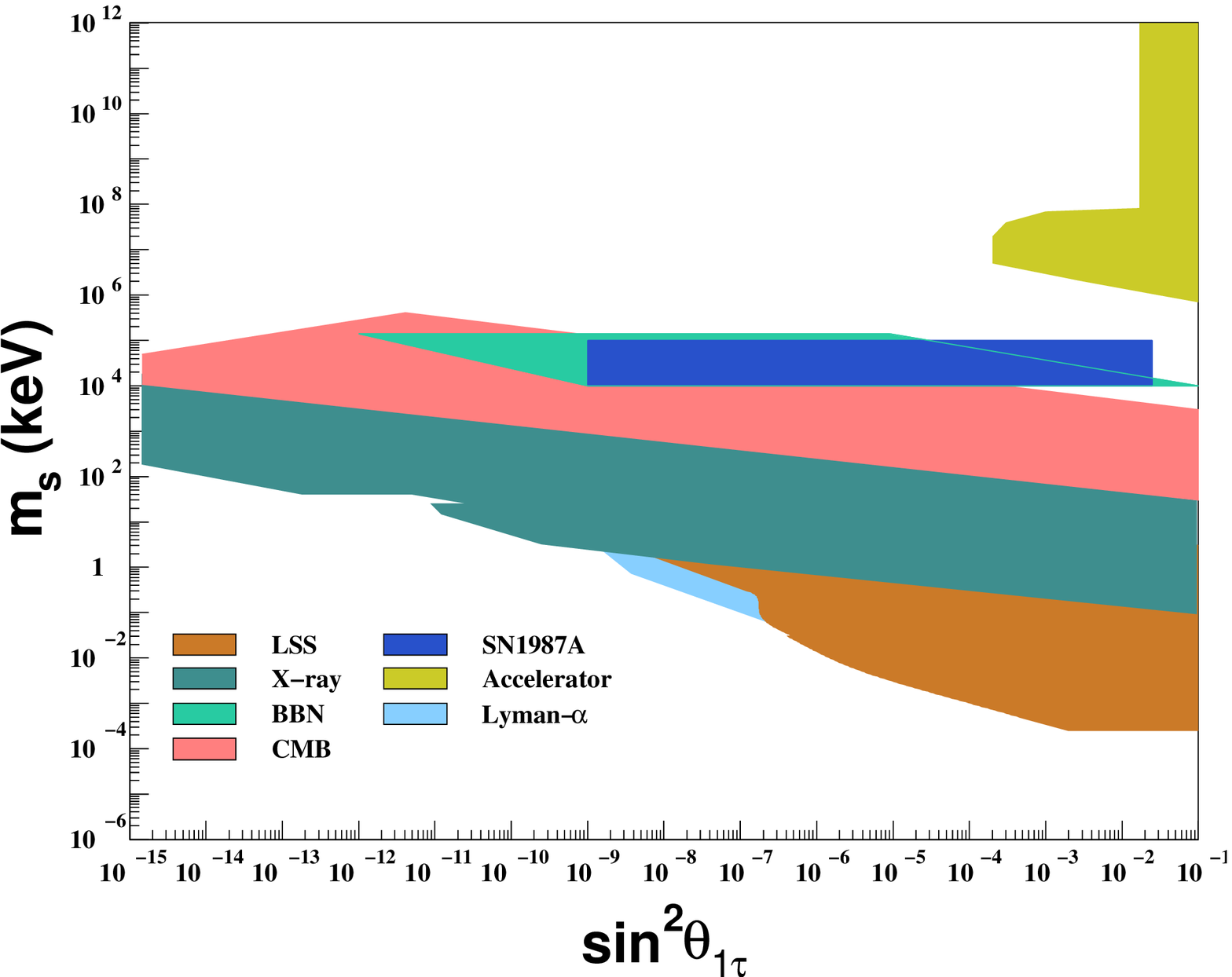}}   
\caption{ Same as in Fig.~\ref{figure:limits_e}, for the sterile neutrino mixing with $\nu_\tau$ only. 
}
\label{figure:limits_tau}
\end{figure}

The current limits on sterile neutrino masses and mixing angles are summarized in Figs.~\ref{figure:limits_e}-\ref{figure:limits_tau}.   The experimental strategy for discovering the sterile neutrinos depends on the mass and mixing parameters.  Many experiments have been proposed and performed over the years.  Accelerator and other laboratory  experiments are able to set limits or discover sterile neutrinos with a large enough mixing angle. The light sterile neutrinos, with masses below $10^2$~eV and large mixing angles, can be discovered in one of the neutrino oscillations experiments~\cite{Smirnov:2006bu}. 

In the eV to MeV mass range, the ``kinks'' in the spectra of beta-decay
electrons can be used to set limits on sterile neutrinos mixed with the
electron neutrinos~\cite{Shrock:1980ct}.  Neutrinoless double beta decays can probe
the Majorana neutrino masses~\cite{Elliott:2002xe}.  An interesting
proposal is to search for sterile neutrinos in beta decays using a complete
kinematic reconstruction of the final state~\cite{Finocchiaro:1992hy,Bezrukov:2006cy}.

For masses in the MeV--GeV range, peak searches in production of neutrinos
provide the limits.  The massive neutrinos $\nu_i$, if they exist, can be 
produced in meson decays, e.g. $\pi^\pm \rightarrow \mu^\pm \nu_i$,  with
probabilities that depend on the mixing in the charged current.  The
energy spectrum of muons in such decays should contain monochromatic
lines~\cite{Shrock:1980ct} at
$ %\begin{equation}
T_i = ( m_\pi^2 + m_\mu^2 - 2 m_\pi m_\mu - m_{\nu_i}^2) / 2 m_\pi. 
% \label{spectrummu}
$ % \end{equation}
Also, for the MeV--GeV masses one can set a number of constraints based on the
decays of the heavy neutrinos into the ``visible'' particles, which would
be observable by various detectors. These limits are discussed in
Ref.~\cite{Kusenko:2004qc}. 

In general, the laboratory experiments, have a limited reach for small mixing angles.  As discussed below, the flavor mixing angles of sterile neutrinos that can be responsible for the pulsar kicks and dark matter are $ \theta \lesssim 10^{-5}$.  The interaction cross sections of such neutrinos are suppresses by factor $ \theta^2 \lesssim 10^{-10}$.  The same factor suppresses the oscillation probability.  Any experiment designed to probe the existence of sterile neutrinos with a small mixing angle (see, e.g.,Refs.~\cite{Finocchiaro:1992hy,Bezrukov:2006cy})  would have to overcome this enormous suppression factor.   The controversial result from  LSND~\cite{Athanassopoulos:1996jb,Athanassopoulos:1996ds,Athanassopoulos:1997er,Athanassopoulos:1997pv}, later refuted by MiniBooNE~\cite{AguilarArevalo:2007it} had affected much of the discussion regarding the sterile neutrinos in the literature.  However, the low-mass, large mixing angle region probed by these experiments was {\em not} motivated at all by particle physics, astrophysics, or cosmology~\cite{Dodelson:2005tp}.  The MiniBooNE results have no bearing on the existence of sterile neutrinos in the mass and mixing ranges in which they can account for the pulsar kicks and dark matter. 

For masses and mixing angles as small as one needs to explain the pulsar kicks and dark matter, the X-ray searches described below probably provide the best opportunity to make a discovery. 

Figs.~\ref{figure:limits_e}--\ref{figure:limits_tau} show an update of exclusion limits discussed in Refs.~\cite{Smirnov:2006bu,Kusenko:2004qc}.   We refer the reader to these papers for a more detailed discussion and a complete list of references.  Here we briefly summarize the origin or various limits shown in the figures.  

For all cosmological limits we assume the minimal possible abundance of sterile neutrinos consistent with standard cosmology, in which the universe was heated to temperatures above 1~GeV.   (See Ref.~\cite{Gelmini:2008fq} for the possibility of very low reheat temperature that could invalidate some of the bounds.)  Some limits on mass and mixing reported in the literature {\em assume} that sterile neutrinos constitute 100\% of cosmological dark matter, even though their abundance would be much lower if no additional production mechanism is introduced besides neutrino oscillations.   We assume the abundance of sterile neutrinos in halos to be at the lowest value consistent with neutrino oscilations.   If some additional production mechanisms are at work (see, e.g., Refs.~\cite{Shaposhnikov:2006xi,Kusenko:2006rh,Petraki:2007gq}), then the X-ray bounds can exclude some additional parameter space as compared to what is shown here.  

The structure formation is affected if sterile neutrinos make a significant
contribution to the mass density of the universe, i.e., if they make up a
non-negligible part of dark matter.  The bounds marked Large Scale Structure
(LSS) are based on the analyzes of Refs.~\cite{Dodelson:2005tp,Smirnov:2006bu}. 
 The Lyman-$\alpha$ constraints shown in
Figs.~\ref{figure:limits_e}--\ref{figure:limits_tau} are based on
Refs.~\cite{Palazzo:2007gz,Boyarsky:2008xj,Boyarsky:2008mt}.  Part of the reason
for the difference with
Refs.~\cite{Narayanan:2000tp,Viel:2005qj,Seljak:2006bg,Viel:2007mv} is that the
authors of Refs.~\cite{Narayanan:2000tp,Viel:2005qj,Seljak:2006bg,Viel:2007mv}
assume that sterile neutrinos produced by some unspecified mechanism account for
the entire observed dark matter, while the authors of
Ref.~\cite{Palazzo:2007gz,Boyarsky:2008xj,Boyarsky:2008mt} consider the effects
of the amounts of sterile neutrinos that can be produced by neutrino
oscillations.  The oscillations produce a calculable amount of sterile neutrinos
in standard cosmology, while some additional mechanisms can produce and ad hoc
amount with a model-dependent free-streaming properties.  The latter is a
possibility, but it is not a good assumption for setting the bounds.  

Likewise, the X-ray limits, based on the non-observation of an X-ray line from the radiative decay of sterile neutrinos (see section~\ref{section:X-ray} for discussion) are shown as in Refs.~\cite{Kusenko:2006rh,Palazzo:2007gz,2008cxo..prop.2676L,2008HEAD...10.2906L}, based on the actual production rates and not on the assumption that all dark matter is accounted for by sterile neutrinos.   This makes the excluded region different from that shown in Refs.~\cite{Abazajian:2001vt,Abazajian:2005gj,Beacom:2005qv,Mapelli:2005hq,Boyarsky:2005us,Abazajian:2006yn,Boyarsky:2006fg,Boyarsky:2006jm,Watson:2006qb,Abazajian:2006jc,Boyarsky:2006ag,Boyarsky:2006hr,RiemerSorensen:2006fh,RiemerSorensen:2006pi,Boyarsky:2006kc,Yuksel:2007xh,Boyarsky:2007ay,RiemerSorensen:2009jp}

The Cosmic Microwave Background (CMB) bounds are shown as in Refs.~\cite{Hannestad:1998zg,Smirnov:2006bu}.  The Big Bang Nucleosynthesis (BBN) constraints are  based on Refs.~\cite{Dolgov:2003sg,Dolgov:2000jw,Smirnov:2006bu}.   The supernova 1987A (SN1987A) bounds are based on Refs.~\cite{1987PASJ...39..521S,Bionta:1987qt,Hirata:1987hu,VanDerVelde:1987hh,Kolb:1988pe,Dolgov:2000jw,Smirnov:2006bu}.  The laboratory bounds, including the accelerator, beta decay, and neutrino oscillations experiments are derived and/or summarized in Refs.~ \cite{Kusenko:2004qc,Smirnov:2006bu}, where the reader can find the detailed discussion and references. 

\section{Dark Matter in the Form of Sterile Neutrinos}
\label{DM_intro}

The singlet fermions are introduced to explain the observed neutrino masses, but the new particles can make up the dark matter.  Because of the small Yukawa couplings, the keV sterile neutrinos are out of equilibrium at high temperatures.  They are not produced in the freeze-out from equilibrium.  However, there are several ways in which the relic population of sterile neutrinos could have been produced.  

\begin{itemize}
 \item Sterile neutrinos could be produced from neutrino oscillations, as 
proposed by Dodelson and Widrow (DW)~\cite{Dodelson:1993je}.  If the lepton asymmetry is
negligible, this scenario appears to be in conflict with a combination of the X-ray
bounds~\cite{Abazajian:2001vt,Abazajian:2005gj,Beacom:2005qv,Mapelli:2005hq,Boyarsky:2005us,Abazajian:2006yn,Boyarsky:2006fg,Boyarsky:2006jm,Watson:2006qb,Abazajian:2006jc,Boyarsky:2006ag,Boyarsky:2006hr,RiemerSorensen:2006fh,RiemerSorensen:2006pi,Boyarsky:2006kc,Boyarsky:2006jm,Yuksel:2007xh,Boyarsky:2007ay,2008cxo..prop.2676L,2008HEAD...10.2906L,Loewenstein:2008yi,RiemerSorensen:2009jp} and the Lyman-$\alpha$
bounds~\cite{Narayanan:2000tp,Viel:2005qj,Seljak:2006qw,Viel:2007mv}.  On the other hand, observations of dwarf spheroids point to a non-negligible free-streaming length for dark
matter~\cite{Kauffmann:1993gv,Hernandez:1998hf,SommerLarsen:1999jx,Klypin:1999uc,Moore:1999nt,Moore:1999wf,Bode:2000gq,Dalcanton:2000hn,Peebles:2001nv,Zentner:2002xt,Simon:2003xu,Gentile:2004tb,Goerdt:2006rw,Strigari:2006ue,Gilmore:2006iy,Wilkinson:2006qq,Boyanovsky:2007zz,Gilmore:2007fy,Wyse:2007zw,Gilmore:2008yp,Koch:2008dc,Munari:2008hb,Siebert:2008uu,Veltz:2008sc}, which
favors warm dark matter, apparently in contradiction with the Lyman-$\alpha$
bounds.  It is also possible that the sterile neutrinos make up only a fraction
of dark matter~\cite{Palazzo:2007gz,Boyarsky:2008xj,Boyarsky:2008mt}, in which
case they can still be responsible for the observed velocities of pulsars~\cite{Kusenko:1997sp,Kusenko:2006rh}.

\item A modification of the DW scenario proposed by Shi and
Fuller~\cite{Shi:1998km} uses a non-zero lepton asymmetry $L$.   The oscillations on 
Mikheev--Smirnov--Wolfenstein (MSW)  resonance~\cite{Wolfenstein:1977ue,Mikheev:1986gs} generate a greater abundance of relic sterile neutrinos with a lower average momentum than in the DW case.  This
results in a colder dark matter with smaller mixing angles, which relaxes the
bounds from small-scale structure and from the X-ray observations.  The
Shi--Fuller (SF) scenario works for a pre-existing lepton asymmetry $L \gtrsim10^{-3}$.  An economical model that can generate the requisite lepton asymmetry was proposed by Laine and
Shaposhnikov~\cite{Laine:2008pg}: decays of the heavier sterile neutrinos could
be responsible for generating the lepton asymmetry of the universe that creates the conditions for producing dark matter in the form of the lighter sterile neutrinos. 

\item The bulk of sterile neutrinos could be produced from decays of gauge-singlet Higgs 
bosons at temperatures above the $S$ boson mass, $T\sim
100$~GeV\cite{Kusenko:2006rh}.  In this case,  the Lyman-$\alpha$ bounds on the
sterile neutrino mass are considerably weaker than in DW case or SF case because the
momenta of the sterile neutrinos are red-shifted as the universe cools down
from $T\sim 100$~GeV~\cite{Kusenko:2006rh,Petraki:2007gq}.

\item Sterile neutrinos could be produced from their coupling to the inflaton~\cite{Shaposhnikov:2006xi} or the radion~\cite{Kadota:2007mv}.  Depending on the time of production, the population of dark-matter particles can be warm or cold.  For example if the mass of the inflaton is below 1~GeV~\cite{Shaposhnikov:2006xi}, one does not expect a significant redshifting of dark matter, which remains warm in this case.   However, if the sterile neutrinos are produced at a higher scale, they can be red-shifted as in the case of the electroweak-scale production~\cite{Kusenko:2006rh,Petraki:2007gq}.  

\end{itemize}

It is important to note that only in the first case, namely, the DW scenario,  the
dark matter abundance is directly related to the mixing angle.  In contrast,
if the relic population of sterile neutrinos arises from the Higgs decays,
their abundance is determined by the coupling $h$ in eq.~(\ref{adding_S}),
while the mixing angle is controlled by a different coupling, as discussed below.  
We also note that both models allow for some production of sterile neutrinos from
oscillations, but in the case of the singlet Higgs decays~(\ref{lagrangianS}), 
the bulk of the sterile dark matter could be produced at  $T\sim 100$~GeV, 
regardless of the value of the mixing angle, which can be vanishingly small.
Nevertheless, the production by oscillations cannot be turned off, and the X-ray
bounds, which depend on the mixing angle, apply even in the case when only a
fraction of dark matter comes from neutrino
oscillations~\cite{Kusenko:2006rh,Palazzo:2007gz,Boyarsky:2008xj,Boyarsky:2008mt}. Of course, 
the number densities of sterile neutrinos in dark halos must be calculated based on the DW production rates.  

Let us now discuss each of these possibilities.  

\subsection{Relic sterile neutrinos produced by neutrino oscillations (Dodelson--Widrow scenario)}

Production of sterile neutrinos by oscillations takes place at relatively low temperatures, below 1~GeV, and depends only on the mass of the sterile neutrino and  
its mixing with the active neutrinos.  It does not depend the high-scale physics that could be
responsible for the neutrino masses.  Other production mechanisms could be active at some higher
temperatures and could produce an additional (usually, colder) component of dark matter, as discussed below.  However, there is no way to turn off the production by oscillations, unless the reheat temperature is very
low~\cite{Gelmini:2004ah,Gelmini:2008fq}.  Any additional high-scale mechanism would lead to a two-component dark matter.  

Dodelson and Widrow~\cite{Dodelson:1993je} proposed that sterile neutrinos produced by oscillations could account for all the dark matter.  The calculation of sterile neutrino abundance was subsequently refined to include various corrections and to properly account for the change in the QCD degrees of freedom that takes place at around the same time~\cite{Abazajian:2001nj,Abazajian:2001vt,Dolgov:2000ew,Abazajian:2005gj,Asaka:2006rw,Asaka:2006nq}.

In general, the singlet neutrino is not an eigenstate of the mass matrix.   One can assume, for example, that the singlet fermion has a non-zero mixing with the electron neutrino, but that the other mixing angles are zero or very small.  Then one finds that the mass 
eigenstates have a simple expression in terms of the weak eigenstates: 
\begin{eqnarray}
| \nu^{\rm (m)}_1 \rangle & = & \cos \theta_m \, | \nu_e \rangle - \sin \theta_m \, |
N_1  \rangle \\ 
| \nu^{\rm (m)}_2 \rangle & = & \sin \theta_m \, | \nu_e \rangle + \cos \theta_m \, |
N_1 \rangle, 
\label{eigenstates}
\end{eqnarray}
where $\theta_m$ is the effective mixing angle, which, at finite temperature and density, could be different from the vacuum mixing angle.  

If the mixing angle $\theta_m $ is small, one of the mass eigenstates, $\nu_1$ behaves very much like a pure $\nu_e$, while the other, $\nu_2$, is practically ``sterile'', {\em i.e.},  its weak interactions are suppressed by a factor $(\sin^2 \theta_m)$ in the cross section.   

Sterile neutrinos are produced through oscillations of active neutrinos.
The relation between their mass and the abundance is very different from
what one usually obtains from freeze-out.  One can trace the production of
sterile neutrinos in plasma by solving the corresponding kinetic equations,  including the effects of QCD degrees or freedom as discussed in Refs.~\cite{Asaka:2006rw,Asaka:2006nq}
Here we will give a very crude but intuitive picture of this production mechanism.  

Neutrino oscillations can convert some of the active neutrinos (in equilibrium) into sterile neutrinos (out of equilibrium).    Matter and the  quantum damping effects inhibit neutrino oscillations~\cite{Stodolsky:1986dx,Barbieri:1989ti,Kainulainen:1990ds,Barbieri:1990vx}.  The mixing of sterile neutrinos with one of the active species in plasma can be represented by an effective, density and temperature dependent  mixing angle~\cite{Dodelson:1993je,Dolgov:2000ew,Abazajian:2001nj}: 
\begin{equation}
\sin^2 2 \theta_m \approx 
\frac{(\Delta m^2 / 2p)^2 \sin^2 2 \theta}{(\Delta m^2 / 2p)^2 \sin^2 
2 \theta + ( \Delta m^2 / 2p \cos 2 \theta - V_m-V_{_T})^2}, 
\label{sin2theta}
\end{equation}
Here $V_m$ and $V_T$ are the effective matter and temperature potentials.
In the limit of small angles and small lepton asymmetry, the mixing angle
can be approximated as 
\beq
\sin  \theta_m \approx
\frac{\sin  \theta}{1+ 0.27 \zeta  \left( \frac{T}{100 \,
\rm MeV} \right)^6 \left( \frac{{\rm keV}^2}{\Delta m^2} \right )
}
\eeq
where $\zeta =1.0$ for mixing with the electron neutrino, and  $\zeta 
=0.30$ for $\nu_\mu$ and $\nu_\tau$.  

Obviously, thermal effects suppress the effective mixing significantly for
temperatures $T> 150 \, (m/{\rm keV})^{1/3}\,$MeV.  If the singlet
neutrinos interact only through mixing, all the interaction rates are
suppressed by the square of the mixing angle, $\sin^2 \theta_m $.  It is
easy to see that these sterile neutrinos are {\em never} in thermal
equilibrium in the early universe.  Thus, in contrast with the case of the
active neutrinos, the relic population of sterile neutrinos is not a result
of a freeze-out\footnote{One immediate consequence of this observation is that the
Gershtein--Zeldovich bound~\cite{Gershtein:1966gg,PhysRevLett.29.669} and the Lee--Weinberg bound~\cite{Lee:1977ua} do not apply to these sterile neutrinos.}.

In the relevant range of parameters, one can roughly approximate the numerical results for the amount of dark matter produced in this scenario~\cite{Dodelson:1993je,Abazajian:2001nj,Abazajian:2001vt,Dolgov:2000ew,Abazajian:2005gj,Asaka:2006nq}:
\beq
\Omega_s \sim 0.2 \left ( \frac{\sin^2 \theta}{3\times 10^{-9}} \right )
\left ( \frac{m_s}{3 \, {\rm keV} } \right )^{1.8} .
\eeq
The range of the masses and mixing angles consistent with dark matter and with the X-ray bounds discussed below forces the mass of the sterile neutrino to be as low as 1-3~keV.  The much improved state-of-the-art calculations~\cite{Asaka:2006rw,Asaka:2006nq} reinforce this conclusion.  However, the Lyman-$\alpha$ bounds~\cite{Narayanan:2000tp,Viel:2005qj,Seljak:2006qw,Abazajian:2006yn,Viel:2007mv} appear to  disfavor this mass range {\em for the production via neutrino oscillations}\footnote{Since different production mechanisms can can generate sterile neutrinos with very different free-streaming properties for the same mass, the mass bounds from Refs.~\cite{Viel:2005qj,Seljak:2006qw,Viel:2007mv} do not apply to models that consider production by other mechanisms, different from Dodelson--Widrow mechanism.  For example, if sterile dark matter is generated at the electroweak scale, the corresponding mass bounds are relaxed by more than factor 3~\cite{Kusenko:2006rh,Petraki:2007gq,Petraki:2008ef,Boyanovsky:2008nc}.}.

\subsection{Lepton asymmetry and the Shi--Fuller scenario}

The production scenario proposed by Dodelson and Widrow~\cite{Dodelson:1993je} is altered  in the presence of a lepton asymmetry of the universe, in which case the production of relic sterile neutrinos can be enhanced by MSW effect~\cite{Wolfenstein:1977ue,Mikheev:1986gs}.  Shi and Fuller~\cite{Shi:1998km} showed that the MSW resonance makes the production more efficient for small missing angles, hence opening up some additional parameter space that is less constrained by the X-ray data.  In addition, the momentum distribution of non-thermal sterile neutrinos produced in this case is colder than in the case of zero lepton asymmetry.~\cite{Shi:1998km,Kishimoto:2006zk,Kishimoto:2008ic}.  This helps ameliorate the tension with the Lyman-$\alpha$ bounds\footnote{See, however, the discussion below regarding some astrophysical evidence in favor of warm dark matter.}.  

For the Shi--Fuller mechanism to work, the lepton asymmetry should be fairly large, several orders of magnitude larger than the baryon asymmetry of the universe.  The lepton asymmetry of the universe today is not known.  Any asymmetry generated at temperatures above the electroweak scale would be distributed in comparable amounts between the baryon ($B$) and lepton ($L$) asymmetries by electroweak sphalerons, which violate $(B+L)$~\cite{Kuzmin:1985mm}.  However, although the standard lore is in favor of much higher temperatures,  there is no direct observational evidence that the temperature of the universe (after inflation) has at any point been higher than the weak scale.   In addition, some lepton number violating processes could generate a sufficiently large lepton asymmetry at a lower temperature.  Affleck--Dine baryogenesis~\cite{Affleck:1984fy,Dine:2003ax,Enqvist:2003gh} can probably produce very different lepton and baryon asymmetries in some cosmological models.  

A very appealing self-contained scenario that allows for the Shi--Fuller production mechanism was proposed by Laine and Shaposhnikov~\cite{Laine:2008pg}.  They considered the most economical model that has enough degrees of freedom to accommodate the active neutrino masses and dark matter in the form of sterile neutrinos.  This model corresponds to $n=3$ in eq.~(\ref{lagrangianM}) and is called $\nu$MSM for \textit{Minimal Standard Model with neutrino masses}~\cite{Asaka:2005an}.  In addition to dark matter, the model is suitable for generating the baryon asymmetry of the universe if the two heavier sterile neutrinos have closely degenerate masses $M_2 \approx M_3 \sim$1-10~GeV (with $ |M_2 - M_3|\sim 1$~keV).  The $\nu$MSM leptogenesis~\cite{Akhmedov:1998qx,Asaka:2005pn} is similar to the widely studied thermal leptogenesis~\cite{Fukugita:1986hr} in that a lepton asymmetry $L$ is generated first, by some processes involving the heavy singlets, and that this lepton asymmetry is then converted into a baryon asymmetry $B\sim L$.  The main difference is that, in thermal leptogenesis, the lepton asymmetry is generated by out-of-equilibrium decays of superheavy sterile neutrinos.   In the case of light sterile neutrinos, the neutrino oscillations replace the decays in generating $L\neq 0$.  Oscillations of active neutrinos into sterile neutrinos leave the in-equilibrium plasma with a non-zero lepton asymmetry.  If the masses of heavy sterile neutrinos, $N_2$ and $N_3$ in eq.~(\ref{lagrangianM}) are nearly degenerate, the CP violation effects due to some complex phases in the neutrino mass matrix are amplified, thus providing for an acceptable baryon asymmetry of the universe, of the order of $10^{-10}$, consistent with the observations.  This scenario is very appealing since no additional ingredients are required, besides those that have been added to the standard model to accommodate the neutrino masses.\footnote{Although a high degree of mass degeneracy among the heavy sterile states is required, it could be a manifestation of some deeper  symmetry~\cite{Shaposhnikov:2006nn}}

However, this is not the end of the story.  At temperatures well below 100~GeV, when the electroweak sphalerons are frozen out, the sterile neutrinos with GeV masses may decay and alter the lepton asymmetry of plasma without changing the baryon asymmetry~\cite{Laine:2008pg}.  If this happens before the universe reaches the temperatures of the order of 100~MeV, the neutrino oscillations responsible for the production of sterile dark matter take place in plasma with a relatively large lepton asymmetry, which is the pre-requisite for  the Shi--Fuller mechanism.  Hence, this very economical model can produce both baryon asymmetry and dark matter.  

Although there is still some tension between what's needed for dark matter and the combined Lyman-$\alpha$ bounds and the X-ray bounds, the Shi--Fuller mechanism helps alleviate this tension by producing colder dark matter (relevant for small-scale structure) for smaller mixing angles (relevant for X-rays). 
An even colder population of dark matter can be produced by some processes at the electroweak scale, as discussed below. 

\subsection{Relic sterile neutrinos produced at the electroweak scale}

As discussed above, the Higgs mechanism can give rise to the neutrino Majorana
mass if the gauge singlet $S$ couples to the neutrinos and acquires a VEV.  It is intriguing that, in this case, the correct dark matter abundance is
obtained for $\langle S  \rangle \sim 10^2$~GeV, suggesting that the singlet may be part of an extended Higgs sector~\cite{Kusenko:2006rh,Petraki:2007gq}.  Conversely, if one chooses the singlet Higgs to have electroweak-scale mass an VEV and if one fixes the coupling as required by dark matter, one predicts the sterile neutrino mass in the range where it needs to be to explain the pulsar kicks.   This
production mechanism is worth examining in more detail because the 
sterile neutrinos produced in this scenario make a colder dark matter than those produced by
oscillations.  The reason is entropy production and red shift.  The number of active degrees of freedom at the electroweak scale is  $g_*(T=100~{\rm GeV})\ge 110.5$, while at the low temperature it reduces to $g_*(T=0.1~{\rm MeV})=3.36$.   The entropy production red-shifts the momenta of the dark matter particles and makes this form of dark matter colder, which is probably in better agreement with observations. 

Let us consider the following Lagrangian: 
\begin{equation} 
{\mathcal L}
   =   {\mathcal L}_{0}+\bar N_{a} \left(i \gamma^\mu \partial_\mu 
\right )
N_{a}  - y_{\alpha a} H \,  \bar L_\alpha N_{a}  - \frac{h_a}{2} \, S \,
\bar {N}_{a}^c N_{a} 
 +V(H,S) + h.c. \,, 
\label{lagrangianS}
\end{equation}
where $ {\mathcal L}_{0}$ includes the gauge and kinetic terms of the Standard
Model, $H$ is the Higgs doublet, $S$ is the real boson which is SU(2)-singlet,
$L_\alpha$ ($\alpha=e,\mu,\tau$) are the lepton doublets, and $ N_{a}$
($a=1,...,n$) are the additional singlet neutrinos.  

The addition of the $SNN$ term opens a new production channel for sterile neutrinos. Dark matter
particles can be produced in decays $S\rightarrow NN $
of $S$ particles in equilibrium.   At later times the sterile neutrinos remain out of equilibrium, while their density and their momenta get red-shifted by the expansion of the universe making dark matter ``colder''.   

Potential $V(H,S)$ yields the mass of $S$ in the minimum: $m_S\sim 10^2$~GeV.  One can estimate the production of  sterile neutrinos by multiplying the $S$ number density (which is $\sim T^3$ for $T>m_S$)  by the $S \rightarrow NN$ decay rate, 
\beq
\Gamma = \frac{f^2}{(16\pi)}m_S,
\label{gamma}
\eeq
and by the time available for the decay, 
$$\tau \sim M_0/T^2,$$
at the latest temperature at which the thermal population of $S$
is still significant, namely 
\beq
T\sim m_S. 
\eeq
At lower temperatures,
the $S$ number density is too small, much smaller than $T^3$. One
obtains an approximate result for the number density of dark matter: 
\beq
 \left( \left. \frac{n_s}{T^3}  \right) \right |_{T\sim m_S}\sim \Gamma \left.
\frac{ M_0}{T^2}\right |_{T\sim m_S} \sim \frac{f^2}{16\pi}
\frac{M_0}{m_S},
\label{approx_n_s}
\eeq
where $M_0 = \left(\frac{45 M_{PL}^2}{4 \pi^3 g_*}\right)^{1/2}  \sim 10^{18} \GeV$
is the reduced Planck Mass.  This approximate result is in good agreement with a more detailed calculation~\cite{Shaposhnikov:2006xi,Petraki:2007gq}.

Once produced, the dark-matter particles remain out of equilibrium.   
Production of sterile neutrinos $\nu_s $ via decay $S \rightarrow NN$ occurs mainly at
temperature $T_{_S}\approx m_{S}$, after which the density of dark matter per
co-moving volume undergoes dilution by some factor $\xi$, and the particle momenta are redshifted by
factor $\xi^{1/3}$.   The value of $\xi$ depends on the number of active light degrees of freedom
at the time of $N$ production. For example, assuming only the degrees of freedom in the  
Lagrangian of eq.~(\ref{lagrangianS}), that is, the Standard Model with the addition of $N$ and $S$ fields, one
obtains 
\beq
\xi =\frac{g_*(T=100\, {\rm GeV})}{g_*(T=0.1\, {\rm MeV})}\approx 33.
\eeq
Based on eq.~(\ref{approx_n_s}) and taking into account the dilution of the dark matter, it is straightforward to show that the $S$ boson decays produce 
\beq
\Omega_{s} \sim 0.2 \left( \frac{f}{10^{-8}}\right)^3  \left(
\frac{\langle S \rangle }{m_{S}}\right) \left( \frac{33}{\xi} \right).
 \label{omega} 
\eeq

On theoretical grounds, it is reasonable to expect a gauge singlet in the Higgs sector; such an extension of the standard model has been the subject of many studies~\cite{McDonald:1993ex,McDonald:1993ey,Vilja:1993uw,Datta:1995qw,Ham:2004cf,Kusenko:2006rh,Ahriche:2007jp,Barger:2006sk,Barger:2007im,Profumo:2007wc,Petraki:2007gq}.  For the electroweak-scale mass and VEV, 
$$ \langle S \rangle \sim m_S \sim 10^2 \, {\rm GeV},$$
the Yukawa coupling $$f\sim 10^{-8}$$ is required to
generate the correct amount of dark matter (see eq.~(\ref{omega})).   So far we just fit one parameter by choosing the
value of another parameter.  However, this value of the Yukawa coupling turns out to be very 
special.  The corresponding Majorana mass $M$ is  
\beq
M_s= f \langle S \rangle \sim 10^{-8} \times (10^2-10^3) \, {\rm GeV} \sim (1-10)\,  {\rm keV}. 
\label{M_s_just_right}
\eeq
This mass is exactly what is needed to explain the pulsar velocities by the anisotropic emission of
sterile neutrinos, as discussed below.   If the value of $\Omega_s $ in eq.(\ref{omega}) was tuned, the mass in 
eq.(\ref{M_s_just_right}) is the outcome of the calculation.  This is intriguing and encouraging.  One can see this as an extra motivation for the search in the relevant mass range.  

Not only is (1--20)~keV mass good for the pulsar kicks, it is also in the most interesting range for
dark matter.  Indeed, if the mass came out to be lower, it would violate the Tremaine--Gunn bound~\cite{Tremaine:1979we} and its generalizations~\cite{Dalcanton:2000hn,Petraki:2008ef,Boyanovsky:2008nc,Boyarsky:2008ju,Gorbunov:2008ka}.  If the mass came out to be much greater, the sterile neutrino would decay too fast to be dark
matter.  Therefore, the mass range   (1--20)~keV is favored by both the pulsar kicks and dark
matter.   We note that theoretical models of neutrino masses can readily produce a sterile neutrino with a required mass~\cite{Farzan:2001cj,Babu:2003is,Babu:2004mj,Sierra:2008wj}.  

The dilution of dark matter by entropy production  affects the average momentum of the
dark-matter particles.  In the case of production at the electroweak scale, the average momentum
of the dark matter particles, relative to the temperature, is 
\beq
\langle p_s \rangle_{(T \ll 1{\rm MeV})} = 0.76 \, T \left [ \frac{110.5}{
g_*(\tilde{m}_{_{S}})
}\right ]^{1/3}.
\label{p_s_redshifted} 
\eeq

We have derived eq.~(\ref{approx_n_s}) and eq.~(\ref{omega}) using some very simple estimates of dark matter production at the electroweak scale.  This admittedly simplistic treatment is in good agreement with the exact solution based on the corresponding kinetic equations~\cite{Shaposhnikov:2006xi,Kusenko:2006rh,Petraki:2007gq}.   The momentum distribution of the resulting dark matter was studied in detail in Refs.~\cite{Petraki:2008ef,Boyanovsky:2008nc}. 

The Lagrangian in eq.(\ref{lagrangianS}) can describe some very interesting Higgs dynamics, which can have a
first-order phase transition~\cite{McDonald:1993ey,Ahriche:2007jp,Profumo:2007wc}.  In this
case the standard electroweak baryogenesis~\cite{Kuzmin:1985mm} can take place in the
course of this first-order phase transition.  The model of eq.~(\ref{lagrangianS}) can
easily be modified to include a sufficient amount of CP violation: all that is
required for a successful baryogenesis is to include the second Higgs
doublet~\cite{McDonald:1993ey}.  

The first-order phase transition can, in principle, further
redshift the dark matter if the dark matter particles are produced mostly before the transition. 
This possibility was analyzed in detail in Ref.~\cite{Petraki:2007gq}.  Finally, one should also consider the possibility of $S$ decays out of equilibrium, which can occur if the doublet--singlet couplings are small.   This possibility was also considered in Ref.~\cite{Petraki:2007gq}; it does not result in a significant production of dark-matter particles for generic ranges of model parameters.

\subsection{How cold is dark matter in the form of sterile neutrinos?}

While the existence of dark matter is well established by a consensus of several independent lines of evidence~\cite{Bertone:2004pz}, the data do not allow to discriminate between a wide variety of candidate particles~\cite{Bertone:2004pz,Murayama:2007ek}.  The list of such particle candidates includes the lightest supersymmetric particles (LSP) and Q-balls predicted by supersymmetry, axions, sterile neutrinos, and a variety of other candidates, well motivated from the theoretical point of view.  It would be a major breakthrough, of course, to obtain some additional information about dark matter directly from astrophysical data.  One possibility to learn about dark matter is to study the small-scale structure, where the particle candidates give very distinct predictions.  

Expressions such as ``cold dark matter'' (CDM) and ``warm dark matter'' (WDM) are often used, but they do not capture the broad variety of small-scale properties exhibited by different dark matter candidates.  A long list of particle candidates~\cite{Bertone:2004pz} labeled traditionally as CDM or WDM fit the observed structure equally well on large scales. However, on small scales, their predictions differ.  The smallest structures in the universe are directly related to particle properties and production history of dark matter.  For instance, if all dark matter is in the form of axions, then the small-scale cutoff in the matter power spectrum is as small as  $10^{-15} M_\oplus$~\cite{Johnson:2008se}.  Supersymmetric dark matter in the form of LSP has a much larger cutoff, $(10^{-6}- 10^2) M_\oplus$~\cite{Profumo:2006bv}, and the dark matter in the form of sterile neutrinos produces an even larger cutoff~\cite{Kusenko:2006rh,Petraki:2007gq,Petraki:2008ef,Boyanovsky:2008nc}, which depends on the mechanism by which these  particles were produced~\cite{Kusenko:2006rh}.  For the sake of unambiguous terminology, one can define CDM as idealized dark matter that has no small-scale suppression in the matter power spectrum, and we define WDM as dark matter that has a non-negligible value of such a cutoff.  A definitive determination of the small-scale power spectrum would be a powerful discriminator between different candidates.  

The data regarding the structures on small scales are becoming available but the
observations are often difficult to interpret.  On the one hand, there are
observations of the Lyman-$\alpha$ forest from gas clouds at high redshifts,
which are interpreted, through modeling, as evidence of small-scale 
structure~\cite{Viel:2005qj,Seljak:2006qw,Strassler:2006ri,Viel:2006kd,Viel:2007mv,Boyarsky:2008xj,Boyarsky:2008mt}.   On the other hand, there are
observations of the motions of stars in dwarf spheroid galaxies, which yield the
gravitational potentials created by dark matter in these dark-matter dominated 
objects~\cite{Hernandez:1998hf,Goerdt:2006rw,Strigari:2006ue,Gilmore:2006iy,Wilkinson:2006qq,Gilmore:2007fy,Wyse:2007zw,Gilmore:2008yp,Koch:2008dc,Munari:2008hb,Siebert:2008uu,Veltz:2008sc}.  These observations are consistent
with cored profiles that are most easily accommodated in the case of warm dark
matter with a suitable small-scale cutoff in the matter power spectrum. 

 Claims have been made that CDM suffers from several inconsistencies between the  N-body simulations and the observations.  These inconsistencies may be resolved by better modeling, but at present they present a challenge.  It is intriguing, however, that WDM is free from many of these problems altogether.  
One challenge to CDM  is a  discrepancy between the number of satellites predicted  in N-body  simulations and the number of observed satellites in galaxies  such
as the Milky Way~\cite{Kauffmann:1993gv,Klypin:1999uc,Moore:1999wf,Moore:1999nt,Willman:2004xc}.
Warm dark matter (WDM) can suppresses small scale structure and dwarf galaxy formation~\cite{Bode:2000gq}, hence ameliorating this discrepancy. 
Several other potential problems in the CDM paradigm may be solved by the reduction of power on small scales due to WDM. For example, WDM can reduce the number of halos in low-density voids bringing it in better agreement with observations~\cite{Peebles:2001nv,Bode:2000gq}. 
Another problem is the relatively low densities  of the galactic cores
implied by the rotation curves as compared to what is predicted from the
$\Lambda$CDM power spectrum~\cite{Dalcanton:2000hn}.  This 
problem can also be rectified by a reduction of the initial power spectrum
of density fluctuations on small scales~\cite{Zentner:2002xt}. 
There appears to be a so called ``angular-momentum'' problem of CDM halos, in which the gas
cools at very early times into small mass halos and leads to massive
low-angular momentum gas cores in galaxies~\cite{SommerLarsen:1999jx}. 
Another apparent inconsistency  is the lack of disk-dominated or pure-disk
galaxies predicted in CDM models~\cite{Governato:2002cv} and the high degree of correlations and organization seen in the statistical properties of galaxies~\cite{2008Natur.455.1082D,GarciaAppadoo:2008fe}.  Both of these inconsistencies can be eliminated if the merger rates were lower than expected in the case of CDM.
All of these inconsistencies arise in CDM models on small scales. In our
opinion, none of these challenges can rule out CDM: a better
understanding of the N-body dynamics, the galaxy formation history and the
possible feedback mechanisms may
provide a resolution for each of these apparent
inconsistencies~\cite{Maccio':2008qt}.   It is true,
however, that warm dark matter in the form of sterile neutrinos is free from all
these small-scale problems altogether, while on large scales WDM fits the data
as well as CDM.  From the particle physics point of view, we see no reason to favor one or the other possibility, because well-motivated candidates exist in both cases.   Improved understanding of structure formation and assembly history in the case of warm dark matter presents a number of interesting observational consequences that can be tested by future observations, such as, for example, the filamentary structure of halos\footnote{This could affect the Lyman-$\alpha$ bounds.} expected from WDM~\cite{Gao:2007yk}.

The momentum distribution of dark matter particles determines how 
warm the dark matter is.  This, in turn, depends on the production mechanism~\cite{Dalcanton:2000hn,Kusenko:2006rh,Petraki:2008ef,Boyanovsky:2008nc}.  A very useful parameter is the phase density that can be defined using the dark matter density $\rho $ and the velocity dispersion $\langle v^2 \rangle$: 
\begin{equation}
 Q= \frac{\rho}{\langle v^2 \rangle ^{3/2}}
\end{equation}
Observations of dwarf spheroidal galaxies can be used to infer bounds on the primordial value of $Q$.  These data have been interpreted as implying a $\sim 1$~keV mass for sterile neutrinos, if these particles make up all of dark matter~\cite{Boyanovsky:2008nc,Wu:2009yr}.

One can also compare different production mechanisms mechanisms using the free-streaming length:  
\beq
 \lambda_{_{FS}} \approx 1 \, {\rm Mpc} \left( \frac{\rm keV}{m_s}
\right) \left(  \frac{\langle p_s \rangle}{3.15 \, T}  \right)_{T\approx {\rm
 1 \, keV}} 
\eeq
The average momentum $\langle p_s \rangle$ depends on the way the sterile dark matter is generated in the early universe: 
\beq
\left(  \frac{\langle p_s \rangle}{3.15 \, T}  \right)_{T\approx {\rm
 1 \, keV}} = 
\left \{ 
\begin{array}{ll}
0.9 &  {\rm for\ production\ off-resonance} \\ 
0.6 & {\rm for\ MSW\ resonance\ (depending\ on}\ L) \\ 
0.2 & {\rm for\ production\ at\ T \gtrsim 100~GeV}
\end{array}
 \right. 
\eeq

This is an admittedly simplistic way to present the results for the small-scale power spectrum.  However, it allows one to juxtapose the three different scenarios and to illustrate their differences.  One can see that the three scenarios produce dark matter that becomes ``warm'' on different length scales.  Hence, one can hope that the future astronomical observations will be able to distinguish between such scenarios.  

The clustering properties of dark matter in the form of sterile neutrinos have been studied in detail  \cite{Abazajian:2005gj,Boyanovsky:2006it,Boyanovsky:2007as,Boyanovsky:2007ay,Boyanovsky:2007ba,Boyanovsky:2007zz,Petraki:2008ef,Boyanovsky:2008he,Boyanovsky:2008nc,Boyarsky:2008ju,Gorbunov:2008ka}, and these studies conclude, in particular, that, if the population of dark-matter sterile neutrinos originate at or above the electroweak scale, as in the model of Refs.~\cite{Kusenko:2006rh,Petraki:2007gq}, then their power spectrum is consistent with the data on small-scale structure, while the other parameters, such as mass and the mixing angle, are within their allowed ranges~\cite{Petraki:2008ef,Boyanovsky:2008he,Boyanovsky:2008nc}.  The phase space arguments~\cite{Tremaine:1979we,Dalcanton:2000hn,Petraki:2008ef,Boyanovsky:2008nc,Boyarsky:2008ju,Gorbunov:2008ka} can be used set the lower bound on the mass of dark-matter sterile neutrinos, which must be heavier than 0.5~keV if they make up all the dark matter and if they were produced at the electroweak scale, with a subsequent cooling, as in Refs.~\cite{Kusenko:2006rh,Petraki:2007gq,Petraki:2008ef,Boyanovsky:2008nc}.

\section{Sterile neutrinos and the supernova explosion}

Sterile neutrinos with masses below a GeV can be produced in supernova explosions.  They can play an important role in the nucleosynthesis~\cite{McLaughlin:1999pd,Fetter:2002xx}, as well as in generating the supernova asymmetries and the pulsar kicks~\cite{Kusenko:1997sp}.  The role played by the sterile neutrinos in the supernova depends on the particle mass.

\subsection{Pulsar velocities}

Radio pulsars are magnetized neutron stars.  They have some very high velocities, the origin of which remains unknown.  The velocities of pulsars are measured either by observation of their angular proper motions~\cite{1982MNRAS.201..503L,1992MNRAS.258..497F,1993MNRAS.261..113H,1994Natur.369..127L}, or by measuring the velocity of an
interstellar scintillation pattern as it sweeps across the
Earth~\cite{1968Natur.218..920S,1970MNRAS.150...67R,1972MNRAS.158..281G,1982Natur.298..825L,1986ApJ...311..183C}.  Each of the two methods has certain advantages.  Using the former method, one can get very precise measurements
with the help of a high-resolution radio interferometer, but such observations take a long time.  Measuring the velocity of a scintillation pattern can be done quickly, but the inference of the actual pulsar velocity must rely on some assumptions about he distribution of scattering material along the line of sight. (For instance, if the density of scatterers is higher near Earth, the pattern moves slower than the pulsar.)  In addition to observational errors, one has to take into account various selection effects.  For example, fast and faint pulsars are under-represented in the data as compared with the slow and bright ones.
Therefore, one has to carefully model the pulsar population to calculate the three-dimensional distribution of pulsar velocities corresponding to the observed two-dimensional projection of their proper  motion~\cite{1997MNRAS.291..569H,1998ApJ...505..315C,Arzoumanian:2001dv,1998ApJ...496..333F}.

Based on the data and population models, the average velocity estimates
range from 250~km/s to 500~km/s~\cite{1968Natur.218..920S,1970MNRAS.150...67R,1972MNRAS.158..281G,1982Natur.298..825L,1986ApJ...311..183C,1997MNRAS.291..569H,1998ApJ...505..315C,Arzoumanian:2001dv}.  The
distribution of velocities is non-gaussian, and there is a substantial
population of pulsars with velocities in excess of 700~km/s.  Some 15\% of
pulsars~\cite{1998ApJ...505..315C,Arzoumanian:2001dv} appear to have velocities greater than 1000~km/s,
while the fastest pulsars have speeds as high as 1600~km/s.  Obviously, an
acceptable mechanism for the pulsar kicks must be able to explain these
very fast moving pulsars.  In addition, one hopes that the pulsar spins can be explained by the same kick mechanism~\cite{Spruit:1998sg}. 

Pulsars are born in supernova explosions, so it would be natural to look
for an explanation in the dynamics of the
supernova~\cite{1970SvA....13..562S,Fryer:2003tc,Scheck:2003rw}.  However,  3-dimensional 
numerical calculations often fail to explain pulsar kicks
in excess of 200 km/s~\cite{Fryer:2003tc}.  The exciting recent discovery of
standing accretion shock instability (SASI)~\cite{Blondin:2002sm,Mezzacappa:2005ju} could lead to
an explanation of both the pulsar kick and the spin. However, it appears that
this mechanism predicts a misalignment between the axis of rotation and the
direction of the pulsar velocity because the spiral SASI 
mode, instrumental in spinning up the pulsar, is expected to give it a kick in the direction orthogonal  to the axis of rotation~\cite{Blondin:2006yw}.  At the same time, the
observational evidence is growing in favor of alignment of the direction of motion with the axis of rotation~\cite{Deshpande:1999qc,Helfand:2000qt,Ng:2003fm,Romani:2004dx,Johnston:2005ka,Wang:2005jg,Kargaltsev:2006py,Ng:2007aw,Ng:2007te}. (As discussed below, such an alignment is a generic prediction of the neutrino driven kick mechanisms.) 

The hydrodynamic kick could be stronger if some large initial asymmetries
developed in the cores of supernova progenitor stars prior to their
collapse.  Goldreich {\em et~al.}~\cite{1997upa..conf..269G} have suggested that unstable
g-modes trapped in the iron core by the convective burning layers and
excited by the $\epsilon$-mechanism may provide the requisite asymmetries.
However, according to numerical calculations~\cite{Murphy:2004ga}, the
$\epsilon$-mechanism may not have enough time to significantly amplify the 
g-modes prior to the collapse.
Evolution of close binaries~\cite{1970ApJ...160L..91G} and asymmetric emission of radio
waves~\cite{1975ApJ...201..447H} have been considered as possible causes of the rapid pulsar
motions.  However, both of these explanations fail to produce a large
enough effect.

Most of the supernova energy, as much as 99\% of the total $10^{53}$~erg
are emitted in neutrinos.  A few per cent anisotropy in the distribution of
these neutrinos would be sufficient to explain the pulsar kicks.
This could be an alternative to a hydrodynamic mechanism which would require a much larger
asymmetry in what remains after the neutrinos are subtracted from the 
energy balance.  The numerical calculations of the supernova assume that
neutrino distribution is isotropic.  What if this is not true?

Since the total energy released in supernova neutrinos is $E\sim 3\times
10^{53}$erg, the outgoing neutrinos carry  the total momentum 
\beq
p_{\nu,{\rm total}} 
\sim 1\times 10^{43} {\rm g\,cm/s}. 
\eeq
A neutron star with mass 1.4$M_\odot$ and speed $v= 1000$~km/s has momentum
\bea
p_* =  (1.4 M_\odot ) v  & \approx & 3 \times 10^{41} \left (
\frac{v}{1000\,   {\rm km/s}} \right ) {\rm g\,cm/s} \nonumber \\ 
& \approx & 0.03 \left ( \frac{v}{1000\, {\rm km/s}} \right ) 
p_{\nu,{\rm total}} . 
\eea
A few per cent asymmetry in the neutrino distribution is, therefore,
sufficient to explain the observed pulsar velocities.  What could cause
such an asymmetry?  The obvious suspect is the magnetic field, which can
break the spherical symmetry and which is known to have an effect on weak
interactions.  We will examine this possibility in detail.  Rather than discussing the calculations presented in the original papers, we will follow a simple (and more pedagogical) discussion from Ref.~\cite{Kusenko:2004mm}.

\subsection{Why a sterile neutrino can give the pulsar a kick}

\label{sec_why}

Active  neutrinos are always {\em produced } with an anisotropy in a strong magnetic field, but they usually {\em escape} isotropically.  The asymmetry in production comes from the asymmetry in the basic weak interactions in the presence of a magnetic field.  Indeed, if the electrons and other fermions are polarized by the magnetic field, the cross sections of the urca processes,
such as $$n+e^+ \rightleftharpoons p+ \bar \nu_e,$$
$$p+e^-\rightleftharpoons n+ \nu_e, $$ depend on the orientation of the
neutrino momentum with respect to the electron spin.  The polarization of electrons creates an asymmetry:
\beq 
\sigma ( \uparrow e^-, \uparrow \nu ) \neq  \sigma ( \uparrow e^-,
\downarrow \nu ) 
\label{sigma_up_down}
\eeq
Depending on the fraction of the electrons in the lowest Landau level, this
asymmetry can be as large as 30\%, which is, seemingly, more than one needs
to explain the pulsar kicks~\cite{1985SvAL...11..123D}.  However, this asymmetry is
completely washed out by scattering of neutrinos on their way out of the
star~\cite{1995ApJ...451..700V,Kusenko:1998yy}.  This is intuitively clear because, as a result of
scatterings, the neutrino momentum is transferred to and shared by the
neutrons.  The neutrinos undergo multiple scattering and remain almost in equilibrium as they diffuse out of the protoneutron star.   In the approximate thermal equilibrium, no asymmetry in the production or scattering amplitudes can result in a macroscopic momentum anisotropy.  This statement can be proved rigorously~\cite{1995ApJ...451..700V,Kusenko:1998yy}.

However, if the neutron star cooling produced some particles whose interactions
with nuclear matter were {\em  even weaker} than those of ordinary
neutrinos, such  particles could escape the star with an anisotropy equal
the anisotropy in their production.  The state $\nu^{\rm (m)}_2$ in equation
(\ref{eigenstates}), whose interactions are suppressed by $(\sin^2 \theta)$
can play such a role.  This could be the explanation of the observed velocities of pulsars.  It is intriguing that the same particle can make up the dark mater.

Let us again consider a model with only one sterile neutrino in which the mass eigenstates are admixtures of active and sterile neutrinos, as in equation (\ref{eigenstates}).
For a sufficiently small mixing angle in matter, $\theta_m$, between $\nu_e$ and
$\nu_s$, only one of the two mass eigenstates, $\nu^{\rm (m)}_1$, is trapped in the
core of a neutron star.  The orthogonal state, $\nu^{\rm (m)}_2$, escapes from the
star freely.  This state is produced in the same basic urca reactions
({\em e.g.}, $n+e^+ \rightleftharpoons p+ \bar \nu_e$ and
$p+e^-\rightleftharpoons n+ \nu_e $) 
with the effective Lagrangian coupling equal the weak coupling
times $\sin \theta_m$.  The production and anisotropy can be greatly enhanced if
the active neutrinos undergo a resonant conversion into the sterile neutrinos
at some density~\cite{Kusenko:1997sp}.  This effect plays an important role in some range of
parameters, although this kind of enhancement is not a necessary condition for the
pulsar kick.  We will consider two ranges of parameters, for which the $\nu_e \rightarrow
\nu_s$ oscillations occur on and off resonance.

\subsection{Pulsar kicks from active--sterile neutrino oscillations} 

Let us consider neutrino cooling during the first 10--15 seconds after the
formation of a hot protoneutron star. 
 Depending on the mass and the mixing angle, there may
or may not be a resonant conversion of the active to sterile neutrinos at
some density in a hot neutron star.  If there is an MSW resonance, the
position of the resonance point depends on the density and the magnetic
field.  The latter introduces the required anisotropy.  In the absence of
the MSW resonance, an off-resonance emission from the entire volume of the
neutron star core is possible.  We will see that this emission is efficient
only after the matter potential has evolved from its initial value to
nearly zero.  This important evolution~\cite{Abazajian:2001nj} requires some time, and can cause a delayed kick~\cite{Kusenko:2008gh}, which, in turn, imposes constraints on the masses and mixing angles.  We
will consider the following three possibilities for the pulsar kick:
\begin{enumerate} 
\item MSW resonance in the core ($\rho > 10^{14}\, {\rm g/cm}^{3}$)
\item MSW resonance outside the core ($\rho < 10^{14}\, {\rm g/cm}^{3}$)
\item an off-resonance emission from the core 
\end{enumerate} 
The three regimes are probably mutually exclusive.  For
example, for all the masses that are consistent with the resonance, the
matter potential evolves very slowly, and there is no significant emission
from the core off-resonance.

\subsection{MSW resonance in the core}
\label{sec_res_core}

For simplicity, let us assume that inside the star there exists a uniform (dipole) magnetic field $\vec{B}$.  Neutrino oscillations in a magnetized medium are described by an effective potential that depends
on the magnetic field~\cite{1989PhRvD..40.1693N,D'Olivo:1989cr,1990PhRvL..64.1088D,Esposito:1995db,1997NuPhB.501...17N,2003PhRvD..68k3003N,Nieves:2003kw,2004PhRvD..70g3001N,Nieves:2004qp} in the following way:
\begin{eqnarray}
V(\nu_{\rm s}) & = & 0,  \label{Vnus} \\
V(\nu_{\rm e})& = & -V(\bar{\nu}_{\rm e}) =  V_0 \: (3 \, Y_e-1+4 \,
Y_{\nu_{\rm e}}), \label{Vnue} \\ 
V(\nu_{\mu,\tau}) & = & -V(\bar{\nu}_{\mu,\tau}) = V_0 \: ( Y_e-1+2 \, 
Y_{\nu_{\rm e}}) \ 
+\frac{e G_{_F}}{\sqrt{2}} \left ( \frac{3 N_e}{\pi^4} 
\right )^{1/3}
\frac{\vec{k} \cdot \vec{B}}{|\vec{k}|}, \label{Vnumu}
\end{eqnarray}
where $Y_e$ ($Y_{\nu_{\rm e}}$) is the ratio of the number density of
electrons (neutrinos) to that of neutrons, $\vec{B}$ is the magnetic field,
$\vec{k}$ is the neutrino momentum, $V_0=10 {\rm eV} (\rho/10^{14}
{\rm g/ cm}^{3} )$.  The magnetic field dependent term in equation
(\ref{Vnumu}) arises from polarization of electrons and {\em not} from
a neutrino magnetic moment, which in the Standard Model is small, and which
we will neglect. (A large neutrino magnetic moment can produce a pulsar
kick through a different mechanism proposed by Voloshin~\cite{1988PhLB..209..360V}.)

The condition for a resonant MSW conversion $\nu_i \leftrightarrow
\nu_j$ is

\begin{equation}
\frac{m_i^2}{2 k} \: \cos \, 2\theta_{ij} + V(\nu_i) = 
\frac{m_j^2}{2 k} \: \cos \, 2\theta_{ij} + V(\nu_j)  
\label{res}
\end{equation}
where $\nu_{i,j}$ can be either a neutrino or an anti-neutrino. 

\begin{figure}[ht]
\centerline{\epsfxsize=5in\epsfbox{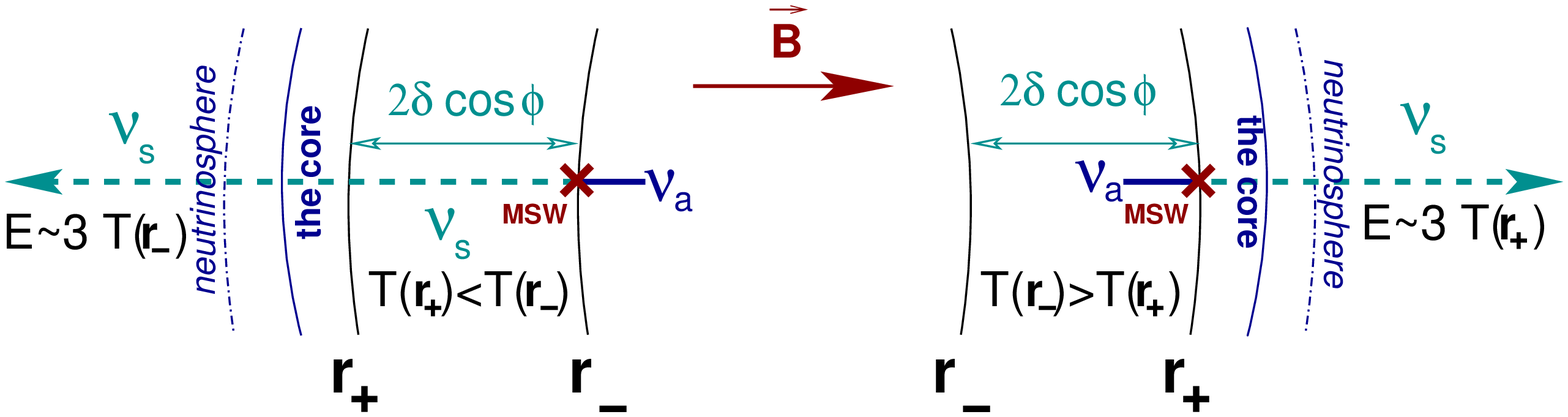}}   
\caption{ For MSW resonance {\em in the core}, the sterile neutrino energy
  depends on the temperature around the resonance point. 
}
\label{figure:core}
\end{figure}

In the presence of the magnetic field, condition (\ref{res}) is
satisfied at different distances $r_\pm$ from the center (Fig.~\ref{figure:core}), depending on the
value of the $(\vec{k} \cdot \vec{B})$ term in (\ref{Vnumu}). The average
momentum carried away by the neutrinos depends on the temperature of the
region from which they escape.  The deeper inside the star, the higher is
the temperature during the neutrino cooling phase.  Therefore, neutrinos
coming out in different directions carry momenta which depend on the
relative orientation of $\vec{k}$ and $\vec{B}$.  This causes the asymmetry
in the momentum distribution.

The surface of the resonance points is 
\begin{equation}
r(\phi) = r_0 + \delta \: \cos \phi, 
\label{delta_r}
\end{equation}
where $\cos \, \phi= (\vec{k} \cdot \vec{B})/(k B)$ and $\delta$ is determined
by the equation 
$(d N_n(r)/dr) \delta \approx 
e \left ( 3 N_e/\pi^4 \right )^{1/3} B$.
This yields 

\begin{equation}
\delta = % \left ( \frac{3 N_e}{\pi^4} \right )^{1/3} \:
%\frac{e}{2} \: B \left / \frac{dN_e(r)}{dr} \right. =
\frac{e \mu_e}{ \pi^2} \: B \left / \frac{dN_n(r)}{dr} \right. ,
\label{delta}
\end{equation}
where $\mu_e \approx (3 \pi^2 N_e)^{1/3} $ is the chemical potential of the
degenerate (relativistic)  electron gas.

In the core of the neutron star, at densities above $10^{14}$~g/cm$^3$, one
can assume the black-body radiation luminosity in sterile neutrinos: 
\beq
F_{\nu_s}({\bf r}) \propto T^4(r). 
\label{black_body}
\eeq 
Then the asymmetry in the momentum distribution is 
\begin{equation}
\frac{\Delta k_s}{k_s} \approx \frac{1}{3}\frac{T^4(r+\delta) -T^4(r-\delta)
}{T^4(r)} 
\approx 
\frac{4}{3} \frac{1}{T} \frac{dT}{dr} (2 \delta), 
\end{equation}
where a factor (1/3) represents the result of integrating 
over angles. 

Now we use the expression for $\delta$ from eq.~(\ref{delta}) and replace
the ratio of derivatives $(\frac{dT}{dr})/(\frac{dN_n}{dr})$ by
$\frac{dT}{dN_n}$: 
\begin{equation}
\frac{\Delta k_s}{k_s} 
\approx \frac{2e}{3 \pi^2} \: \left (
\frac{\mu_e}{T} \frac{dT}{dN_n} \right) B.
\label{dk1}
\end{equation}
To calculate the derivative in (\ref{dk1}), we assume approximate thermal
equilibrium.  Then one can use the relation between the
density and the temperature of a non-relativistic Fermi gas: 
\begin{equation}
N_n=\frac{2(m_n T)^{3/2}}{\sqrt{2} \pi^2}
\int \frac{\sqrt{z} dz}{e^{z-\mu_n/T}+1}, 
\label{fermi}
\end{equation}
where $m_n$ and $\mu_n$ are the neutron mass and chemical potential. 
The derivative $(dT/dN_n)$ can be computed from (\ref{fermi}).  Finally,
\begin{equation}
\frac{\Delta k_s}{k_s} = 
\frac{8 e\sqrt{2}}{\pi^2} \: 
\frac{\mu_e \mu_n^{1/2}}{m_n^{3/2}T^2} \ B. 
\end{equation}
We have assumed that only one of the neutrino species undergoes a resonance
transition into a sterile neutrino.  The energy, however, is shared between
6 species of active neutrinos and antineutrinos.  Therefore, the final
asymmetry due to anisotropic emission of sterile neutrinos is 6 times
smaller: 
\bea
\frac{\Delta k_s}{k} & = & \frac{1}{6}\frac{\Delta k_s}{k_s} =
\frac{4 e\sqrt{2}}{3 \pi^2} \: 
\frac{\mu_e \mu_n^{1/2}}{m_n^{3/2}T^2} \ B = \nonumber \\ & = & 
0.01 
\left ( \frac{\mu_e}{ 100 \rm \,MeV} \right )
\left ( \frac{\mu_n}{ 80 \rm \,MeV} \right )^{1/2}
\left ( \frac{20 \rm \,MeV}{T} \right )^2
\left ( \frac{B}{ 3\times 10^{16} {\rm \,G} }\right )
\label{dk2}
\eea
This estimate~\cite{Kusenko:1997sp} can be improved by considering a more detailed
model for the neutrino transport and by taking into account time evolution
of chemical potentials discussed below.  However, it is clear that the
magnetic field inside the neutron star should be of the order of
$10^{16}$~G.  The approximation used in equation (\ref{black_body}) holds
as long as the resonant transition occurs deep in the core, at density of
order $10^{14} \, {\rm g\,cm^{-3}}$.  This, in turn, means that the sterile
neutrino mass must be in the keV range.

\subsection{Resonance at densities below $ 10^{14}$g/cm$^3$}
\label{sec_res_lessdense}

For smaller masses, the resonance occurs at smaller densities.  Outside the
core, fewer neutrinos are produced, while there is a flux of
neutrinos diffusing out of the core.  Therefore, the approximation
(\ref{black_body}) is not valid.

\begin{figure}[ht]
\centerline{\epsfxsize=5in\epsfbox{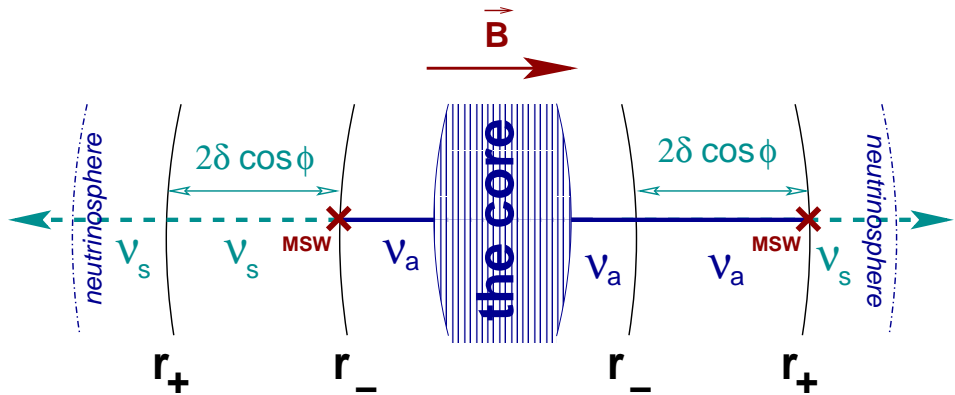}}   
\caption{ For MSW resonance {\em outside the core}, the neutrino passes
between $r_-=r_0-\delta \cos \phi$ and $r_+=r_0+\delta \cos \phi$ as a
sterile $\nu_s$ on one side of the star, while it still propagates as an 
active $\nu_a$ on the other side.  The active neutrinos interact and
deposit some extra momentum on the right-hand side, 
between $r_-$ and $r_+$.  Since the neutron star is a gravitationally bound
object, the momentum deposited asymmetrically in its outer layers gives the
whole star a kick. }
\label{figure:layer}
\end{figure}

Outside the core, the active neutrinos can interact
with matter and deposit momentum to the neutron star medium.  After an
active neutrino is converted into a sterile neutrino, it no longer
interacts with matter and comes out of the star.  Some of the electron
neutrinos are absorbed on their passage through a layer in the neutron star
atmosphere.  In particular, there is a 
charged-current process 
\begin{equation}
\nu_e n \rightarrow e^- p^+, 
\end{equation}
which has a cross section $\sigma = 1.8 \: G_{_F}^2 E_\nu^2 $, where
$E_\nu$ is the neutrino energy.  If the resonant conversion
$\nu_e\rightarrow \nu_s$ occurs at different depths for different
directions, the neutrinos may spend more time as active on one side of the
star than on the other side of the star, as shown in
Fig.~\ref{figure:layer}.  Hence, they deposit more momentum through their
interactions with matter on one side than on the other side.  Let us
estimate this difference.  (This argument was used before in application to
active neutrino transport~\cite{Kusenko:1998bk}.  Here we adopt it to sterile
neutrinos.)

Depending on the magnetic field, the resonance lies at different depths,
eq.~(\ref{delta_r}).  Hence, the neutrinos on one side of the star pass an
extra layer of thickness $(2 \delta \cos\phi)$ as active, while the
neutrinos on the other side pass this layer as sterile.  The active
neutrinos going through a layer of nuclear matter with thickness $2\delta$
have an extra probability
$$P_\delta = (2\delta)\, \sigma \, N_n$$ 
to interact and deposit momentum $k\sim E_\nu$ to the neutron star. This
momentum is not balanced by the neutrinos on the other side of the star
because they go through the this layer as sterile neutrinos.  The cross
section $\sigma \approx 1.8 G_{F}^2 E^2$. 

Obviously, the neutron star as a whole is a gravitationally bound object,
so any momentum deposited on one side of the star gives the whole neutron
star a kick.  
%The anisotropic momentum deposition into a layer of a neutron
%star outside the core is illustrated in Fig.~\ref{figure:layer}. 
%\footnote{Apparently, this needs to be emphasized because this
%  points has been misunderstood by some otherwise sensible
%astrophysicists, whole criticized this argument.} 

The difference in the momentum deposition per electron neutrino between
the directions $\phi$ and $-\phi$ is
%
%\begin{eqnarray}
\beq
\frac{\Delta k_s}{k}  =  (2 \delta \cos \phi) \: 
N_n \: \sigma 
 =  1.8 \: G_{_F}^2 E_\nu^2 \: \frac{\mu_e}{Y_e} \: \frac{eB}{\pi^2} 
\: h_{N_e} \:  \cos \phi,
\eeq
%\end{eqnarray}
where we have used eq.~(\ref{delta}) and introduced the scale height 
of the electron density $h_{N_e}=[d(\ln N_e)/dr]^{-1}$.

We take $Y_e\approx 0.1$, $E_\nu \approx 3 T\approx 10 $~MeV, $\mu_e\approx
50$~MeV, and $h_{N_e}\approx 6$~km.  We assume $T\approx 3$~MeV because it
is a realistic average temperature around the neutrinosphere, in agreement
with theoretical models as well as observations of the supernova
SN1987A~\cite{1987PASJ...39..521S}. (This temperature is lower than the core
temperature used earlier.)

%we obtain $\Delta  
%k_e/E_{\nu_e} = 0.01 (B/2 \ {\rm MeV}^2) \cos \phi $.     
After integrating over angles and taking into account that only one
neutrino species undergoes the conversion, we obtain the final result for the
asymmetry in the momentum deposited by the neutrinos.  

\begin{equation}
\frac{\Delta k_s}{k} = 0.03 
\left ( \frac{T}{3\, \rm MeV} \right )^2
\left ( \frac{\mu}{50 \, \rm MeV} \right )
\left ( \frac{h}{6\, \rm km} \right )
\left ( \frac{B}{10^{15} \, {\rm G}} \right ),
\label{final} 
\end{equation}
This is, clearly, a sufficient asymmetry for the pulsar kick. The
corresponding region of parameters is shown as region ``2'' in
Fig.~\ref{fig:range}.
 
The above estimates are valid as long as $\delta$ is much smaller than the
mean free path.  One can also describe the propagation of neutrinos in this
region using the so called diffusion approximation.  It was
used for the neutrino transport by Schinder and Shapiro~\cite{1982ApJ...259..311S} in
planar approximation, and it can be applied to the pulsar kicks~\cite{Kusenko:1998bk,Barkovich:2001rp,Barkovich:2002wh,Barkovich:2004jp}.

\subsection{Off-resonance transitions} 
\label{sec_offres}

Let us now consider the case of the off-resonance emission from the core.
For masses of a few keV, the resonant condition is not satisfied
anywhere in the core.  In this case, however, the off-resonant
production of sterile neutrinos in the core can occur through ordinary urca
processes.  A weak-eigenstate neutrino has a $\sin^2\theta $ admixture
of a heavy mass eigenstate $\nu_2$.  Hence, these heavy neutrinos can be
produced in weak processes with a cross section suppressed by $\sin^2\theta
$. 

Of course, the mixing angle in matter $\theta_m$ is not the same as it is
in vacuum, and initially $\sin^2\theta_m \ll \sin^2\theta$.  However, as
Abazajian et al.~\cite{Abazajian:2001nj} have pointed out, in the presence
of sterile neutrinos the mixing angle in matter quickly evolves toward its
vacuum value.  When $\sin^2\theta_m \approx \sin^2\theta$, the production
of sterile neutrinos is no longer suppressed, and they can take a fraction
of energy out of a neutron star.  

Following Abazajian, Fuller, and Patel,\cite{Abazajian:2001nj} one can estimate the
time it takes for the matter potential to evolve to zero from its initial
value $V^{(0)}(\nu_e)\simeq (-0.2 ...+ 0.5) V_0$.  The time scale for this
change to occur through neutrino oscillations off-resonance is    
\begin{eqnarray}
\label{timeeqoffres}
 \tau_{_V}^\mathrm{off-res}  & \simeq &
\frac{4 \sqrt{2} \pi^2 m_n}{G_{\!\!_F}^3 \rho}
\frac{ (V^{(0)}(\nu_e))^3}{(\Delta m^2)^2 \sin^2 2 \theta } \frac{1}{\mu^3}
\\
& \sim & 
 \frac{6 \times 10^{-9} s}{\sin^2 2 \theta} 
\left (\frac{V^{(0)}(\nu_e)}{0.1 \mathrm{eV}} \right )^3 
\left (\frac{50 \mathrm{MeV}}{\mu} \right )^3 \left ( \frac{ 
10 \mathrm{keV}^2
}{\Delta m^2
} \right )^2. \nonumber   
\label{timeeqoffresnumerical}
\end{eqnarray}

As long as this time is much smaller than 10 seconds, the mixing angle in
matter approaches its vacuum value in time for the sterile neutrinos to
take out some fraction of energy from a cooling neutron star.  

The urca processes produce ordinary neutrinos with some asymmetry depending
on the magnetic field~\cite{1985SvAL...11..123D}.  The same asymmetry is present in the
production cross sections of sterile neutrinos.  However, unlike the active
neutrinos, sterile neutrinos escape from the star without rescattering.
Therefore, the asymmetry in their emission is not washed out as it is in
the case of the active neutrinos~\cite{1995ApJ...451..700V,Kusenko:1998yy}.   Instead, the asymmetry in
emission is equal the asymmetry in production.
 
The number
of neutrinos $dN$ emitted into a solid angle $d\Omega $ can be written as
\begin{equation}
\frac{dN}{d\Omega}= N_0(1+ \epsilon \cos \Theta_\nu ), 
\end{equation}
where $\Theta_\nu$ is the angle between the direction of the magnetic field
and the neutrino momentum, and $N_0$ is some normalization factor.  The
asymmetry parameter $\epsilon$ is equal
\begin{equation}
\epsilon = \frac{g_{_V}^2-g_{_A}^2}{g_{_V}^2+3g_{_A}^2} 
k_0   
\left ( \frac{  E_{\rm s}}{ E_{\rm tot}} \right ),  
\end{equation} 
where $g_{_V}$ and $g_{_A}$ are the axial and vector couplings, ${ E_{\rm
tot}}$ and $ { E_{\rm s}} $ are the total neutrino energy and the energy
emitted in sterile neutrinos, respectively.  The number of electrons in the
lowest Landau level, $k_0$, depends on the magnetic field and the chemical
potential $\mu$, but it can be in the  $0.1 < k_0 < 0.7$ range for some realistic values of these parameters~\cite{Fuller:2003gy}.

\begin{figure}[ht]
%\epsfxsize = 8cm  
%\epsfbox{k0.eps}
\centerline{\epsfxsize=4.1in\epsfbox{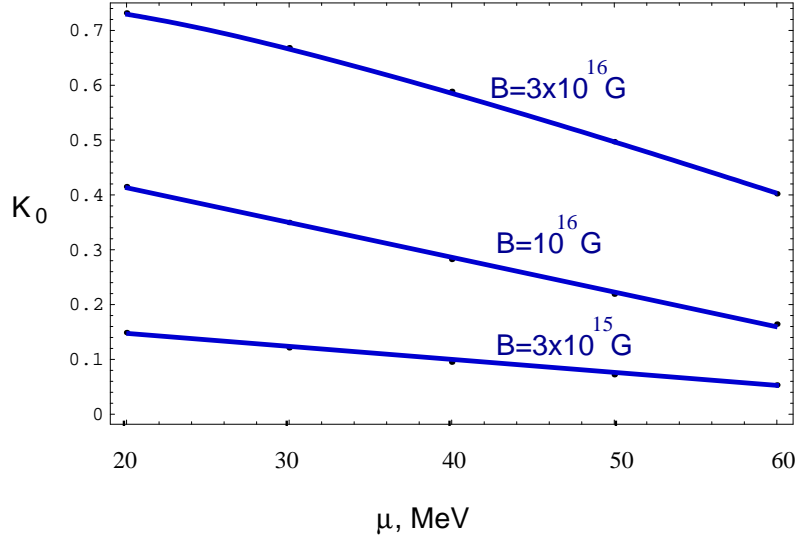}}   
\caption{The fraction of electrons in the lowest Landau level as a function
chemical potential.  The value of the magnetic field is shown next to each curve~\cite{Fuller:2003gy}.  }
\label{figure:asymmT20}
\end{figure}

The momentum asymmetry in the neutrino emission is 
\begin{equation}
\epsilon \sim 0.02 
\left ( \frac{k_0}{0.3}\right ) 
\left ( \frac{ r_{_E}}{0.5} \right), 
\label{epsilon_final} 
\end{equation}  
where $r_{_E}$ is the fraction of energy carried by the sterile neutrinos.
To satisfy the constraint based on the observation of neutrinos from
supernova SN1987A, we require that $r_{_E}<0.7$.  As can be seen from Fig.~\ref{figure:asymmT20}, 
the asymmetry in equation (\ref{epsilon_final})
can be of the order of a few per cent, as required, for magnetic fields
$10^{15}-10^{16}$~G.

Surface magnetic fields of ordinary radio pulsars are estimated to be of
the order of $10^{12}-10^{13}$G.  However, the magnetic field inside a
neutron star may be much higher~\cite{1983MNRAS.204.1025B,1992ApJ...392L...9D,1999ApJ...510L.115K}, probably up to
$10^{16}$G.  The existence of such a strong magnetic field is suggested by
the dynamics of formation of the neutron stars, as well as by the stability
of the poloidal magnetic field outside the pulsar~\cite{1992ApJ...392L...9D}.  Moreover, the
discovery of soft gamma repeaters and their identification as
magnetars~\cite{1992ApJ...392L...9D,1999ApJ...510L.115K}, {\em i.e.}, neutron stars with {\em surface}
magnetic fields as large as $10^{15}$~G, gives one a strong reason to
believe that the interiors of many neutron stars may have magnetic fields
as large as $10^{15}-10^{16}$~G.  There are also plausible
physical mechanisms that can generate such a large magnetic field
inside a cooling neutron star~\cite{1992ApJ...392L...9D}.   The  $\alpha-\Omega$ dynamo mechanism makes the fields grow during the first 10-15 seconds after the formation of the neutron star~\cite{1992ApJ...392L...9D}.  The corresponding equations can be analyzed in linear regime, until the perturbations grow large.  The saturation field of the order of $10^{16}$ G, at which the linear analysis is no longer valid is probably a good estimate of the interior field.

\subsection{Pulsar kicks from the active neutrinos alone?} 

One can ask whether the sterile neutrino is necessary and whether the
oscillations of active neutrinos alone could explain the pulsar kicks.  The
interactions of muon and tau neutrinos in nuclear matter are characterized
by a smaller cross section than those of the electron neutrinos.  This is
because the electron neutrinos $\nu_e$ interact through both charged and
neutral currents with electrons, while $\nu_\mu$ and $\nu_\tau$ interact
with electrons 
through neutral currents alone.  Therefore, nuclear matter is less 
transparent to $\nu_e$ than to $\nu_{\mu,\tau}$, $\bar{\nu}_{\mu,\tau}$.
As a result, the surface of last scattering for $\nu_{\mu,\tau}$ and
$\bar{\nu}_{\mu,\tau}$ lies (about a kilometer) deeper than that of
$\nu_e$.  The electron antineutrino can interact through charged currents
with {\em positrons} while they are present in nuclear matter.  The
$\bar{\nu}_e$ mean free path starts out closer to that of $\nu_e$, but, as
the number of positrons diminishes during the cooling period, this mean
free path increases and becomes comparable to that of $\nu_{\mu,\tau}$ and
$\bar{\nu}_{\mu,\tau}$.

Since the $\mu-$ and $\tau-$neutrinospheres lie inside the electron
neutrinosphere, it is possible that neutrino oscillations could convert a
$\nu_e$ into $\nu_\mu$ or $\nu_\tau$ at some point between the two
neutrinospheres, where the $\nu_e$ is {\em trapped}, but the
$\nu_{\mu,\tau}$ is free-streaming.  Then the shift in position of the MSW
resonance would result in an anisotropy of the outgoing momentum.  This
mechanism could explain the pulsar kicks, but it would require one of the
active neutrino masses to be of order $10^2$~eV~\cite{Kusenko:1996sr}.  This is not
consistent with the present experimental data on neutrino masses.

\subsection{What if neutrinos have a large magnetic moment?} 

The neutrino magnetic moment in the Standard Model is very small, $\mu_\nu
\approx 3\times 10^{-19}(m_\nu/{\rm eV})\mu_{_B} $, where $\mu_{_B}
=e/2m_e$ and $m_\nu$ is the (Dirac) neutrino mass.  This is why we have so
far neglected any effects of direct neutrino interactions with the magnetic
field.  

However, the present experimental bounds allow the neutrino magnetic
moments to be as large as $10^{-12}\mu_{_B} $.  If, due to some new
physics, the neutrino magnetic moment is large, it may open new
possibilities for the pulsar kick.  Voloshin~\cite{1988PhLB..209..360V} has proposed an
explanation of the pulsar kick based on the resonant spin-flip and
conversion of the left-handed neutrinos into the right-handed neutrinos,
which then come out of the neutron star.  Voloshin argued that the magnetic
field inside a neutron star may be irregular and may have some
asymmetrically distributed ``windows'', through which the right-handed
neutrinos could escape.  The resulting asymmetry may, indeed, explain the 
pulsar velocities.

Other kinds of new physics may cause the pulsar kicks as well~\cite{Horvat:1998st,Barkovich:2001rp,Farzan:2005yp}.

\subsection{Time delays}

\begin{figure}
\epsfxsize=9cm 
\centerline{\epsfbox{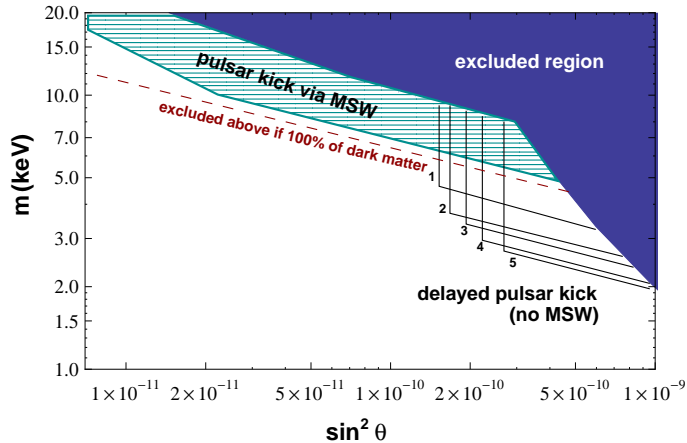}}
\caption{ The allowed regions for delayed kicks with delays from 1 through 5 seconds (assuming the
 other parameters are fixed) are shown by
black solid lines marked by the numbers representing the delay time in seconds~\cite{Kusenko:2008gh}.  
} 
\label{figure:delays}
\end{figure}

As one can see from eq.~(\ref{timeeqoffres}), there can be a non-negligible time delay prior to the onset of an efficient emission of sterile neutrinos.  The matter potential relaxes to zero on some time scales related to the mass and mixing parameters of the sterile neutrinos~\cite{Kusenko:2008gh}.  After the equilibrium is achieved at $V_m \approx 0, $  the effective mixing angle in matter is close to that in vacuum, and the emission of sterile neutrinos proceeds at a much higher rate.   There is a considerable uncertainty in the equilibration time given by  eq.~(\ref{timeeqoffres}) because several parameters are functions of time and position in the star and are not known precisely.  In addition, the equilibration does not have to occur simultaneously in the entire star.  Nevertheless, the very possibility of the delay and its connection with the fundamental neutrino parameters can allow one to distinguish this mechanism from some alternatives.  The studies of the pulsar populations can eventually determine the delay based on the observational parameters~\cite{Ng:2007aw}.   The allowed ranges of parameters corresponding to some time delays are shown in Fig.~\ref{figure:delays}. 

\subsection{{$B-v$} correlation?} 

The neutrino kick mechanism does not predict a correlation
between the direction of the surface magnetic fields and the pulsar
velocity.  The kick velocity is determined by the magnetic field {\em
inside} the hot neutron star during the {\em first seconds} after the
supernova collapse.  Astronomical observations, such as measurements of the pulsar spindown rate,  can be used to infer the {\em surface} magnetic fields of pulsars some {\em millions of years}
later.  The relation between the two is highly non-trivial because of the
complex evolution the magnetic field undergoes in a cooling 
neutron star.  Let us outline some stages of this evolution.

Immediately after the formation of the hot neutron star the magnetic field
is expected to grow due to differential rotation, thermal
effects~\cite{1983MNRAS.204.1025B}, and convection~\cite{zeldovich}. The dynamo effect
can probably account for the growth of the magnetic field to about
$10^{15}-10^{16}$~G~\cite{1992ApJ...392L...9D}.

The growth of the magnetic field takes place during the first ten seconds
after the supernova collapse, in part because the neutrino cooling causes
convection.  At the same time, during the neutrino cooling phase, the
neutron star receives a kick.  The magnetic field relevant for the kick is
the average dipole component of the magnetic field inside the neutron star during the first 10 seconds of explosion.  There is no reason to believe that it has the same direction or magnitude as the surface field at the end of the neutrino cooling phase.  This, however, is
only one of several stages in the evolution of the magnetic field.  Next,
at some temperature below $0.5$~MeV, the nuclear matter becomes a type-II
superconductor.  The magnetic field lines form the flux tubes, reconnect, and
migrate.  Next, over some millions of years, the pulsar rotation converts the
magnetic field energy into radio waves and causes the field to evolve even
further.  The end result of this evolution is, of course, a configuration
of magnetic fields that is very different from what it was five seconds
after the onset of the supernova.

Clearly, the magnetic field inside a hot young neutron star is not expected
to have much correlation with the surface field of a present-day pulsar.

\subsection{Spin-kick from neutrinos} 

Spruit and Phinney~\cite{Spruit:1998sg} pointed out that the pulsar rotational
velocities may also be explained by the kick received by the neutron stars
at birth.  The core of the progenitor star is likely to co-rotate with the
whole star until about 10 year before the collapse.  This is because the
core should be tied to the rest of the star by the magnetic field.
However, then the angular momentum of the core at the time of collapse is
$10^3$ times smaller than the angular momentum of a typical pulsar.  Spruit
and Phinney~\cite{Spruit:1998sg} have pointed out that the kick that accelerates the pulsar can
also spin it up, unless the kick force is exerted exactly head-on.

The neutrino kick can be strongly off-centered, depending on the
configuration of the magnetic field.  If the magnetic field of a pulsar is
offset from the center, so is the force exerted on the pulsar by the
anisotropic emission of neutrinos.  This mechanism may explain
simultaneously the high spatial velocities and the unusually high rotation
speeds of nascent neutron stars.\cite{1997MNRAS.291..569H}  It was suggested by Phinney (private communication) that a highly off-centered magnetic field could be generated by a {\it thin-shell} dynamo in a hot
neutron star.  Since the neutron star is cooled from the outside, a
convective zone forms near the surface and, at later times, extends to the
interior.  While convection takes place in the spherical shell, the dynamo
effect can cause a growth in the magnetic field.
Thin-shell dynamos are believed to be responsible for generating magnetic
fields of Uranus and Neptune~\cite{1991Icar...93...82R,1993JGR....9818659C,1996JGR...101.2177H}.
According to the {\em Voyager~2} measurements~\cite{1986Sci...233...85N}, the magnetic fields of both Uranus and Neptune are off-centered dipoles,  tilted with respect to the axis of rotation.  Unlike other planets, which have convection in the deep interior and end up with a well-centered dipole 
field, Uranus and Neptune have thin spherical convective zones near the
surface, which explains the peculiarity of their dynamos.  During the
first few seconds after the supernova collapse, convection in a neutron
star also takes place in a spherical layer near the surface.  The
thin-shell dynamo can, in principle, generate an off-center magnetic field
in a neutron star, just like it does in Uranus and
Neptune.  As a result, the anisotropic neutrino emission can give the pulsar the kick, and also spin it up. 

\subsection{$\Omega -v$ correlation}

\begin{figure}[ht]
\epsfxsize=10cm   %width of figure - will enlarge/reduce the figures
% \epsfbox{fig3.eps}
% \figurebox{2cm}{3cm}{} %to have a box alone 
\centerline{\epsfxsize=2.0in\epsfbox{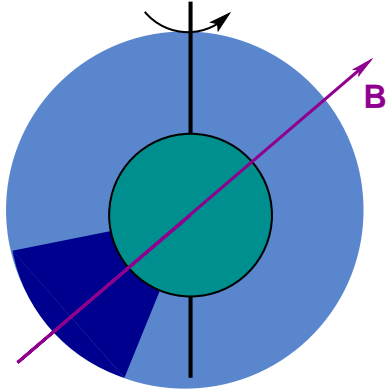}}   
\caption{
The anisotropy in outgoing neutrinos, shown here by the density in shading, creates a thrust that  averages to zero along any horizontal axis, because of the rotation of the protoneutron star about the vertical axis.  The vertical component of the force, aligned with the axis of rotation, is not affected by the rotation.  The pulsar velocity should be aligned with the surviving (vertical) component of the neutrino-driven thrust.   Hence, one expects a directional correlation between the axis of rotation and the direction of pulsar motion.  Such a correlation has already been observed~\cite{Deshpande:1999qc,Helfand:2000qt,Ng:2003fm,Romani:2004dx,Johnston:2005ka,Wang:2005jg,Kargaltsev:2006py,Ng:2007aw,Ng:2007te}.  The rapid motion of the higher-density jet can also be the source of gravitational radiation that can be detected in the event of a nearby supernova~\cite{Loveridge:2003fy}.
}
\label{figure:Omega_v}
\end{figure}

While the $B-v$ correlation is not expected from a neutrino-driven mechanism, one does expect the correlation between the direction of the pulsar motion and the axis of rotation.  This is because the neutrino anisotropy axis is associated with the dipole magnetic field, which is rotating with the star.  If the pulsar is accelerated by anisotropic emission of neutrinos over the time period of several seconds, the components of the thrust orthogonal to the axis of rotation would average to zero, while the component along the axis would not be affected by rotation (see Fig.~\ref{figure:Omega_v}).  As a result, one expects the pulsar to receive a non-vanishing kick along its rotational axis.  To probe this correlation, one needs to perform accurate measurements of the polarization of the radio signal from the pulsar.   The recent observations confirm the $\Omega -v$ correlation~\cite{Deshpande:1999qc,Johnston:2005ka,Wang:2005jg}.   In addition, in several cases {\em Chandra} images allow one to identify the rotational axis of the pulsar with the observed symmetry axis of the pulsar wind nebula.   These observations also confirm the predicted $\Omega -v$ correlation~\cite{Ng:2003fm,Ng:2007aw,Ng:2007te}.   

While the observed correlation between the direction of rotation and the pulsar
velocity does not provide a definitive prove of the mechanism, it was a generic
prediction of the neutrino-driven mechanism which turned out to agree with the
data.  

\subsection{Gravity waves from a neutrino-driven pulsar kick} 

In the event of a nearby supernova, the neutrino kick can produce gravity
waves that could be detected by LIGO and LISA~\cite{Loveridge:2003fy,MosqueraCuesta:2004bc}.
These gravity waves can be produced in several ways.  

Obviously, the departure from spherical symmetry is a necessary condition
for generating the gravity waves.  A neutron star being accelerated by
neutrinos is not moving fast enough to generate gravitational waves from it
own motion.  However, the anisotropy in the outgoing neutrinos turns out to
be sufficient to produce an observable signal in the event of a nearby
supernova. 

\begin{figure}[ht]
%\epsfxsize=10cm   %width of figure - will enlarge/reduce the figures
%\epsfbox{fig3.eps}
%\figurebox{2cm}{3cm}{} %to have a box alone 
\centerline{\epsfxsize=5.2in\epsfbox{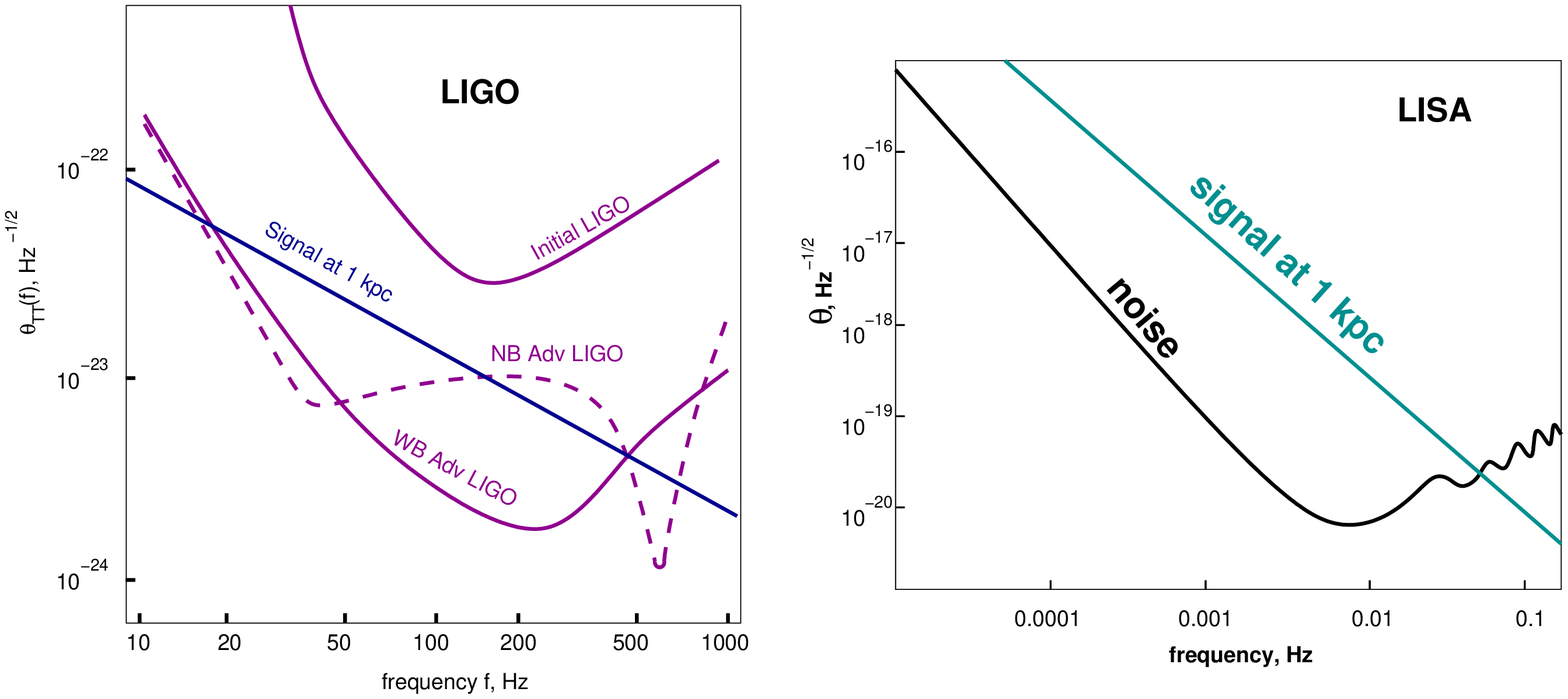}}   
\caption{
Gravity waves signal at LIGO and LISA frequencies calculated by 
Loveridge~\cite{Loveridge:2003fy}.
 }
 \label{figure:grav_waves}
\end{figure}

Most of the neutrinos come out isotropically and can be neglected.
However, a few per cent of asymmetrically emitted neutrinos move along the
direction of the magnetic field.  In general, the magnetic field is not
aligned with the axis of rotation (Fig.~\ref{figure:Omega_v}), and, therefore, the outgoing neutrinos
create a non-isotropic source for the waves of gravity. (Water jet produced
by a revolving lawn sprinkler is probably a good analogy for the geometry
of this source.)  The  signal was calculated by Loveridge.\cite{Loveridge:2003fy} It can be observed by advanced LIGO or LISA
if a supernova occurs nearby, as shown in Fig.~\ref{figure:grav_waves}.
In addition, the neutrino conversion itself may cause gravity waves
coming out of the core~\cite{MosqueraCuesta:2000qh,2002PhRvD..65f1503C,MosqueraCuesta:2000qk,MosqueraCuesta:2004bc}.

The gravitational waves signal from a nearby supernova can be observed by Advanced LIGO or LISA.  If a fast pulsar emerges from the supernova remnant but the signal is not seen by the gravitational waves detectors, it would imply that either the pulsar kick mechanism has nothing to do with asymmetric emission of neutrino, or that the magnetic field is aligned with the axis of rotation.  The radio observations of the nascent pulsar could help determine whether the latter is the case.  Obviously, the gravitational waves from a nearby supernova would be extremely helpful in determining whether the pulsar kick is neutrino-driven or not.  

\subsection{Supernova asymmetries}

The supernova explosion can be altered by sterile neutrinos in several ways, in particular, via the motion of the neutron star accelerated by the emission of sterile neutrinos.  Such a motion breaks the spherical symmetry and introduces an additional  reason for asphericity that could be seen in supernova remnants, as well as optical observations of distant explosions using  spectropolarimetry and other methods~\cite{2005AstL...31..245L,Leonard:2006qy,Maeda:2008mw,Tanaka:2008bs,Tanaka:2008bt}.  In addition, the overall energy deposition behind the shock can increase as a result of convection facilitated by the motion of the neutron star~\cite{Fryer:2005sz}. 

\begin{figure}[ht]
\centerline{\epsfysize 7cm  \epsfbox{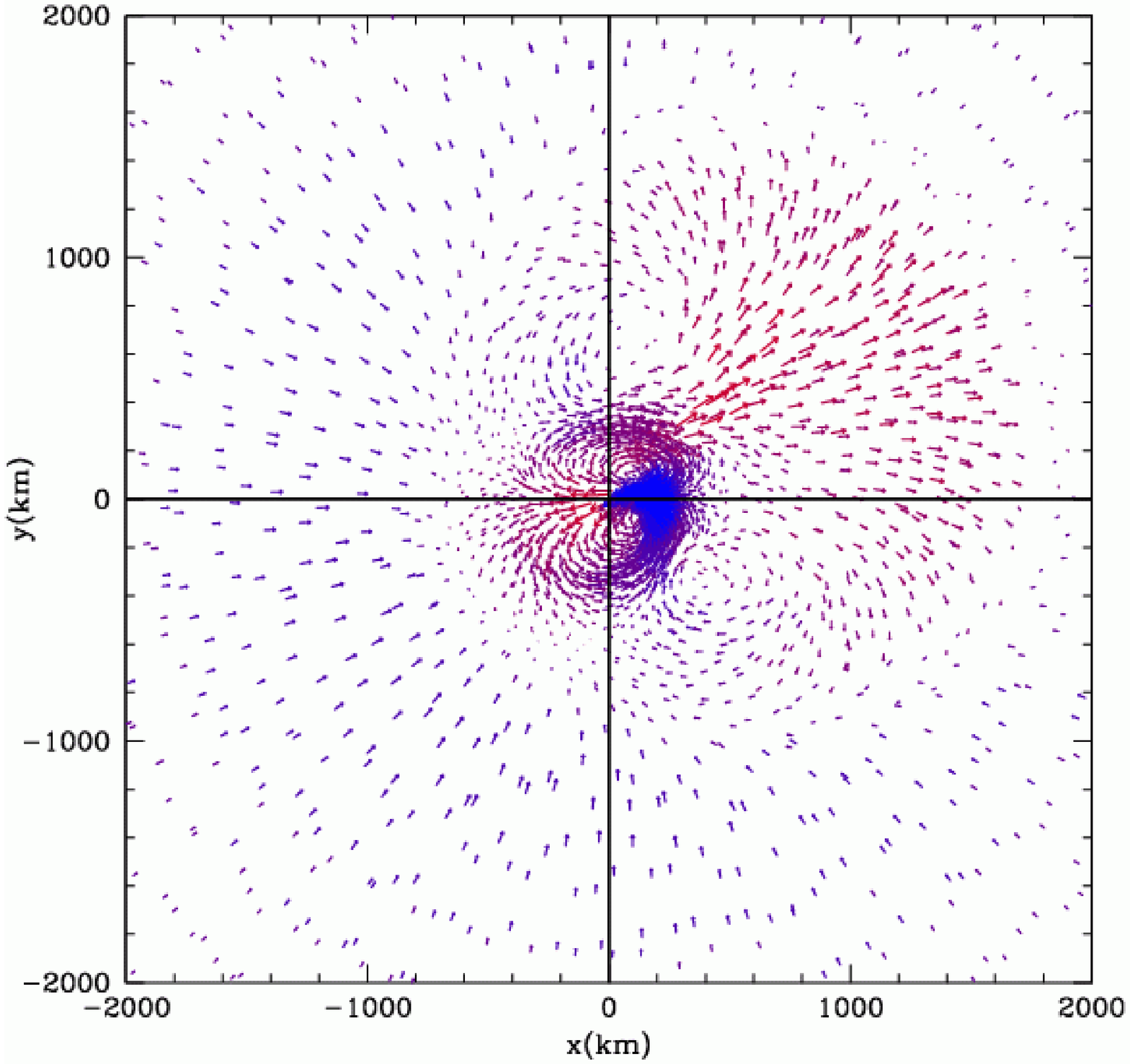} \epsfysize 7cm \epsfbox{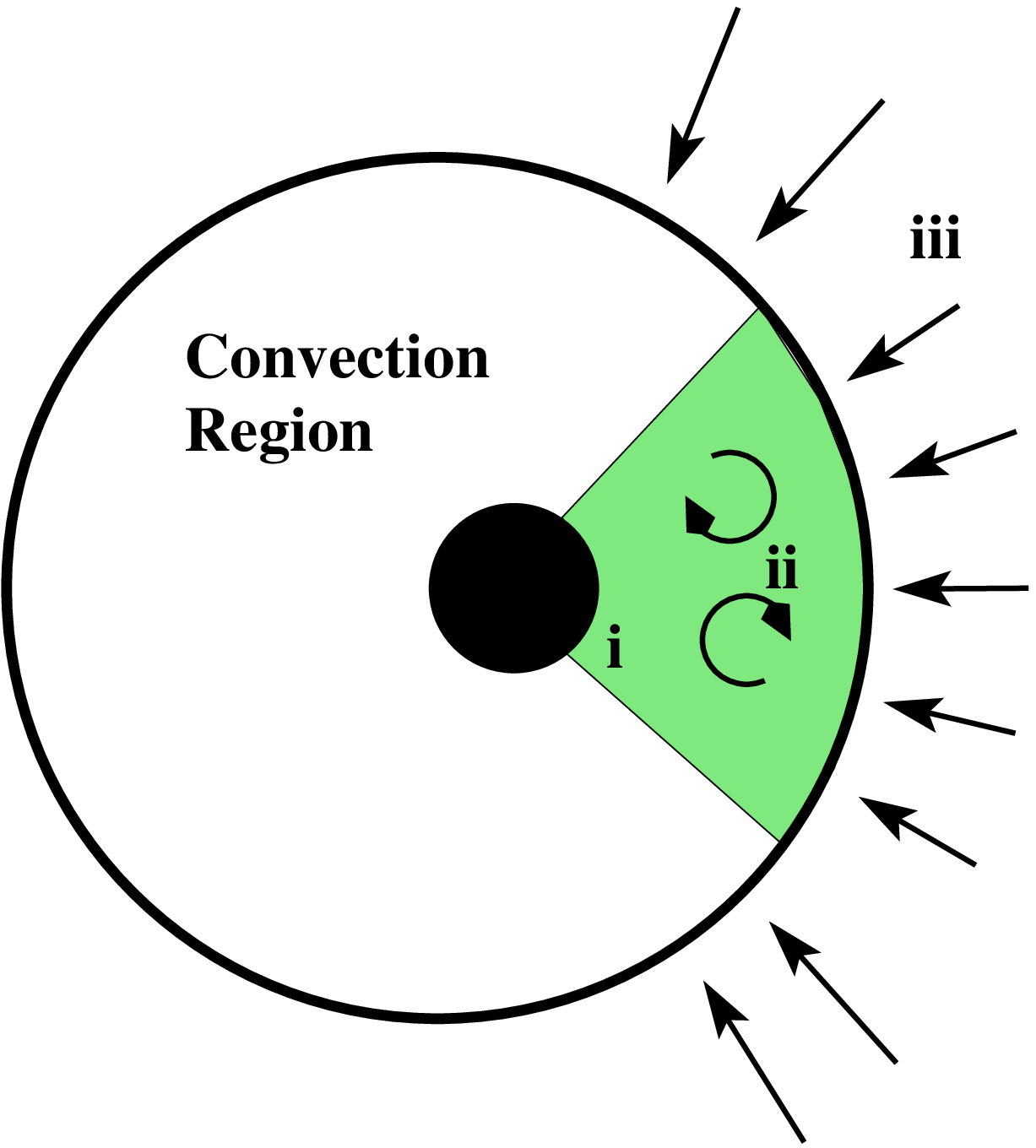}}
\caption{The results of numerical calculations~\cite{Fryer:2005sz} show the increase in convection in front of a neutron star moving from left to right due to a neutrino-driven kick.  The enhanced convection puts more energy behind the shock, which leads to a stronger explosion.   Another prediction of the neutrino-driven kick is the formation of asymmetric jets, with a stronger jet in the direction of he neutron star motion~\cite{Fryer:2005sz}.  This is different from hydrodynamic mechanisms, in which one expects the stronger jet to point in the opposite direction (due to momentum conservation).  Future observations may help distinguish between these mechanisms.}
\label{figure:Fryer}
\end{figure}

Numerical simulations of the neutrino-driven kick mechanism show that the motion of the neutron star can seed and drive convection in front of the moving star, adding the energy to the ejecta and ultimately helping the explosion~\cite{Fryer:2005sz}.  The resultant explosion is asymmetric, with the strongest ejecta motion roughly in the direction of the neutron star kick (Fig.~\ref{figure:Fryer}).  There should be more mixing in the direction of the neutron star motion, and, in particular, nickel should mix further out in this direction~\cite{2005ApJ...635..487H}.  Such an extended mixing in the direction of the neutron star kick has already been observed for the fast moving compact star in Sgr A East~\cite{2005ApJ...631..964P}.  The asymmetry with an enhancement in the same direction as the neutron star motion is in sharp contrast with the consequences of ejecta-driven mechanisms, which predict the motion of the ejecta in the opposite direction, in accordance with momentum conservation.  If both the neutron star and the ejecta receive some additional momentum in the same direction, the compensating momentum can be carried by sterile neutrinos.  

The difference between the codirectional asymmetry in the case of a neutrino driven kick and counter-directional asymmetry expected from ejecta-driven mechanisms can be used to distinguish between the two classes of mechanisms based on the observations of the supernova remnants~\cite{Fryer:2005sz}.  

\subsection{Sterile neutrinos with the $\sim \,$keV masses and the energy transport}
\begin{figure}[ht]
\centerline{\epsfxsize=3in\epsfbox{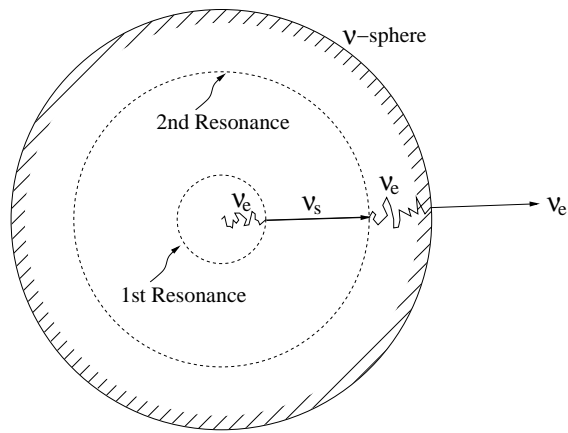}}   
\caption{ The double MSW resonance in the core can facilitate  energy transfer out of the core to outer regions and can have a dramatic effect on the explosion~\cite{Hidaka:2006sg,Hidaka:2007se} 
}
\label{figure:hidaka_fuller}
\end{figure}

Since the sterile neutrinos interact with nuclear matter very weakly, they can be very efficient
at transporting the heat in the cooling proto-neutron star, altering the 
dynamics of the supernova~\cite{Hidaka:2006sg,Hidaka:2007se,Fuller:2008rh}.  
In particular, sterile neutrinos with masses in the keV range could undergo resonant MSW conversion at some density deep inside the core, below the neutrinosphere.   This is very different from the active neutrinos, which cannot have an MSW resonance at these high densities because their mass differences are too small.  The energy transfer inside the supernova can change dramatically because of the resonant conversion followed by a non-diffusive leap, followed by another conversion, as shown in Fig.~\ref{figure:hidaka_fuller}.    
This could lead to an enhancement of the overall supernova explosion. 

An additional enhancement can come from the increase in convection in front of the neutron star propelled by the asymmetric emission of sterile neutrinos~\cite{Fryer:2005sz}.  Naively one could expect that, since sterile neutrinos carry away some energy, the overall energy of supernova should be smaller.  However, this effect is over-compensated by the enhanced convection in front of the moving neutron star, so that the resulting energy of the shock is higher than in the case without sterile neutrinos~\cite{Fryer:2005sz}.

\section{Heavy sterile neutrinos and their effect on the supernova explosions}

A substantial range of parameters with large masses and mixing angles (Figs.\ref{figure:limits_e}-\ref{figure:limits_tau}) is ruled out by the combination of the laboratory experiments and cosmology~\cite{Kusenko:2004qc,Smirnov:2006bu}.  However, there are still  windows in which the sterile neutrinos are not only allowed, but also desirable because they could  augment core collapse supernova shock energies by enhancing energy transport from the core to the vicinity of the shock front~\cite{Falk:1978kf,Zatsepin:1978ac,Fuller:2008rh}.   

One example is a sterile neutrino with mass between 145 and 250 MeV.  This range of masses is particularly interesting from the point of view of its induced contribution to the mass matrix of active neutrinos~\cite{Smirnov:2006bu}.   In this mass range the sterile neutrinos  decay predominantly into a pion and a light fermion. If the mixing angle with the
electron neutrino is negligible, and the sterile neutrino mass $M_s$ is in the range $m_{\pi^0} <
M_s < (m_{\pi^0}+m_\mu)$, the daughter pion is the neutral pion, which decays into  two photons: 
$\nu^{\rm (s)} \rightarrow \nu^{\rm (a)} \pi^0 \rightarrow \nu^{\rm (a)} \gamma
\gamma$~\cite{Dolgov:2000jw}, where $a=(\mu,\tau) $.  This decay mode, with lifetime $\sim 0.1\,{\rm
s}$, changes the impact of sterile neutrinos on the supernova  explosion. To distinguish sterile
neutrinos that decay mainly into photons from the other types, the authors of Ref.~\cite{Fuller:2008rh} called them {\em
eosphoric}.\footnote{From the ancient Greek god $E \omega\sigma\varphi \acute{o} \rho o \varsigma$,
the bearer of light.} 

While not in contradiction with supernova 1987A bounds~\cite{Kolb:1988pe}, the decays of these neutrinos could produce a flux of energetic active neutrinos, detectable by future neutrino observations in the event of a galactic supernova~\cite{Fuller:2008rh}.  Moreover, the relevant range of sterile neutrino masses and mixing angles can be probed in future laboratory experiments~\cite{Bernardi:1985ny,Bernardi:1987ek,Baranov:1992vq,Nedelec:2000rz,Astier:2001ck}.

\section{X-ray Detection of Relic Sterile Neutrinos}
\label{section:X-ray}

\begin{figure}[ht]
%\epsfxsize = 8cm  
%\epsfbox{k0.eps}
\centerline{\epsfxsize=5in\epsfbox{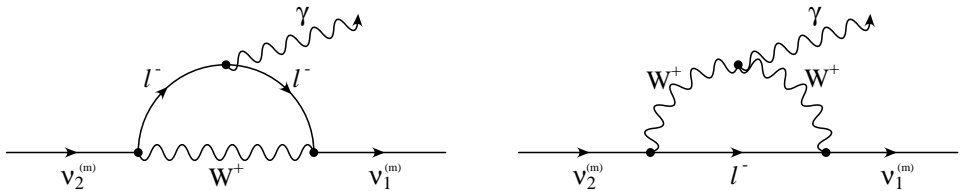}}   
\caption{
Radiative decay of sterile neutrinos, $\nu^{\rm (m)}_2\rightarrow\gamma \nu^{\rm (m)}_1$.  The
X-rays produced by these decays can be detected by the X-ray telescopes,
such as {\em Chandra}, {\em Suzaku},  {\em XMM-Newton}, and the future {\em
  Constellation-X}.  
}
\label{figure:one_loop}
\end{figure}

The main decay mode of sterile neutrinos in the keV mass range is $\nu_s\rightarrow 3\nu$.  This decay mode is ``invisible'' due to the low energy of the daughter neutrinos.  In addition to this leading mode of decay that occurs through a tree-level diagram, there are also  one-loop diagrams (Fig.~\ref{figure:one_loop}) that allow for a photon in the final state.  
Therefore, the sterile neutrinos can decay into the lighter neutrinos and an the
X-ray photons: $\nu_s\rightarrow\gamma \nu_a$~\cite{Pal:1981rm}.   The radiative decay width is equal to~\cite{Pal:1981rm,Barger:1995ty} 
\begin{eqnarray}
 \Gamma_{\nu_s\rightarrow\gamma \nu_a} &=& \frac{9}{256\pi^4}\,  \alpha_{\rm EM}\, {\rm G}_{\rm F}^2 \, \sin^2 \theta \, m_{\rm s}^5 \nonumber \\ 
& = & \frac{1}{1.8\times 10^{21} {\rm s}}\  \sin^2 \theta \ \left( \frac{m_{\rm s}}{\rm keV}\right)^5, 
\end{eqnarray}
and the corresponding lifetime is many orders of magnitude longer than the age of the universe.   However, since sterile neutrinos are produced in the early universe by neutrino oscillations and, possibly, by other mechanisms as well, every dark matter halo should contain some fraction of these particles.   Given a large number of particles in these astrophysical systems, even a small decay width can make them observable via the photons produced in the radiative decay.  This offers, arguably, the best opportunity to detect these particles. 
Since $\nu^{\rm (m)}_2\rightarrow\gamma \nu^{\rm (m)}_1$ is a  two-body decay, the resulting photons have energy $$E_\gamma = m_{\rm s}/2,$$ which corresponds to a line broadened only by the velocity dispersion of the dark matter particles in a given halo.  This line, with photon energy of a few keV, can be observed using an X-ray
telescope~\cite{Abazajian:2001vt}.  

A broad range of astrophysical systems can provide suitable targets for such observations.  A concise discussion and comparison of such observational targets can be found in Ref.~\cite{Boyarsky:2006fg}.  First, there should be a signal from distant unresolved sources, in the form of isotropic extragalactic background with a (red-shifted) photon line.  Second, closer objects with high densities of dark matter provide an even better target: the Milky Way halo, the clusters of galaxies, and the dwarf spheroid galaxies can give different contributions depending on the field of view of one's instrument.   In choosing the best target for a specific instrument, one usually looks for the most dark matter to fit within the field of view (FOV).  This is because the contribution of each individual source scales with distance $r$ as $1/r^2$, while the fraction of volume slice of thickness $\delta r$ at some fixed distance $r$ that fits within the FOV scales as $r^2$.  Therefore, to first approximation, it is the total amount of dark matter in the FOV that determines the sensitivity of dark matter search.   Based on this reasoning, as well as the considerations of observational and instrumental backgrounds, dwarf spheroid galaxies are probably best suited for searching sterile neutrinos with small-FOV instruments, such as Suzaku X-ray  telescope~\cite{2008cxo..prop.2676L,2008HEAD...10.2906L,Boyarsky:2006fg}.  

We note in passing that the search strategy for decaying dark matter is quite different from the searches carried out for annihilating dark matter, such as supersymmetric neutralinos, for example.   The annihilation rates, and, therefore, the expected signals from annihilating dark matter are proportional to the square of the density.  Such predictions are sensitive to the matter power spectrum on small scales and the morphology of such a signal is difficult to disentangle from the point sources.   In contrast, sterile neutrino decays are expected to produce the flux proportional to the first power of dark matter number density, not the density squared.  This fact, combined with the lack of small structure on sub-kpc scales for WDM, renders the theoretical predictions for Suzaku~\cite{2008cxo..prop.2676L,2008HEAD...10.2906L} and other searches fairly robust.

% \subsection{Different types of constraints}

Several types of constraints have been reported in the literature, and, depending on the underlying assumptions, the excluded regions may differ.  The X-ray flux depends on the sterile neutrino abundance.   In view of the variety of possible production mechanisms, one can considers two types of X-ray limits that give an answer to one of the following questions: 

\begin{itemize}
 \item Can a sterile neutrino with a given mass and mixing angle exist (even if it contributes only a small fraction to dark matter)? 

\item Can a sterile neutrino with a given mass and mixing account for all of dark matter?

\end{itemize}

Both questions can be addressed with the use of the X-ray data, but the
resulting bounds on the mass and the mixing angle can differ significantly.  The
mass and mixing angle alone do not determine the abundance of relic sterile
neutrinos.  They determine unambiguously the amount of relic neutrinos produced
by active-to-sterile neutrino oscillations~\cite{Dodelson:1993je}, but this need
not be the only source of sterile neutrinos.   Some additional amounts could be
produced from the Higgs decays~\cite{Kusenko:2006rh,Petraki:2007gq} or the
inflaton~\cite{Shaposhnikov:2006xi} or radion~\cite{Kadota:2007mv} decays, or
any other mechanism, is controlled by some additional parameters, unrelated to
the mixing angle.  It is true, however, that, regardless of how many sterile
neutrinos were produced at high temperature, the low-temperature production by
oscillations cannot be switched off (except in the low-reheat cosmological
scenarios~\cite{Gelmini:2004ah}).  So, for every mass and mixing angle one can
calculate the minimal sterile neutrino abundance and determine the number of
sterile neutrinos in galactic halos using the methods developed in
Refs.~\cite{Kainulainen:1990ds,Dodelson:1993je,Shi:1998km,Abazajian:2001nj,Abazajian:2002yz,Dolgov:2000ew,Kishimoto:2006zk,Asaka:2006rw,Asaka:2006nq,Abazajian:2008dz,Laine:2008pg}.  The abundance calculated this way gives the
lower bound on the sterile neutrino abundance.  If this quantity exceeds the
amount of dark matter determined from cosmological observations, as would be the
case for large mixing angles, then the particle is ruled out.  Otherwise,  one
can use the X-ray observations to set a model-independent bound on the mass and
mixing angle for the particle that can exist but may contribute only a small
fraction of the observed dark
matter~\cite{Kusenko:2006rh,Palazzo:2007gz,Boyarsky:2008xj,Boyarsky:2008mt}. 
Such a particle may not solve the missing matter problem, but it can be
responsible for the pulsar kicks in some range of parameters. 

Alternatively, one ask whether a sterile neutrino with a given mass and mixing can account for all the dark matter, regardless of how it is produced in the early universe.  In this case, one does not have to calculate the amount produced by neutrino oscillations: the relic abundance is assumed to be that of dark matter.  This, in turn, determines the flux of X-rays expected from a given astrophysical object, for example, a dwarf spheroid galaxy.  The X-ray bounds obtained this way will, in general, be stronger that those based on production by oscillations 

\begin{figure}[ht!]
  \centering
  \includegraphics[width=14cm]{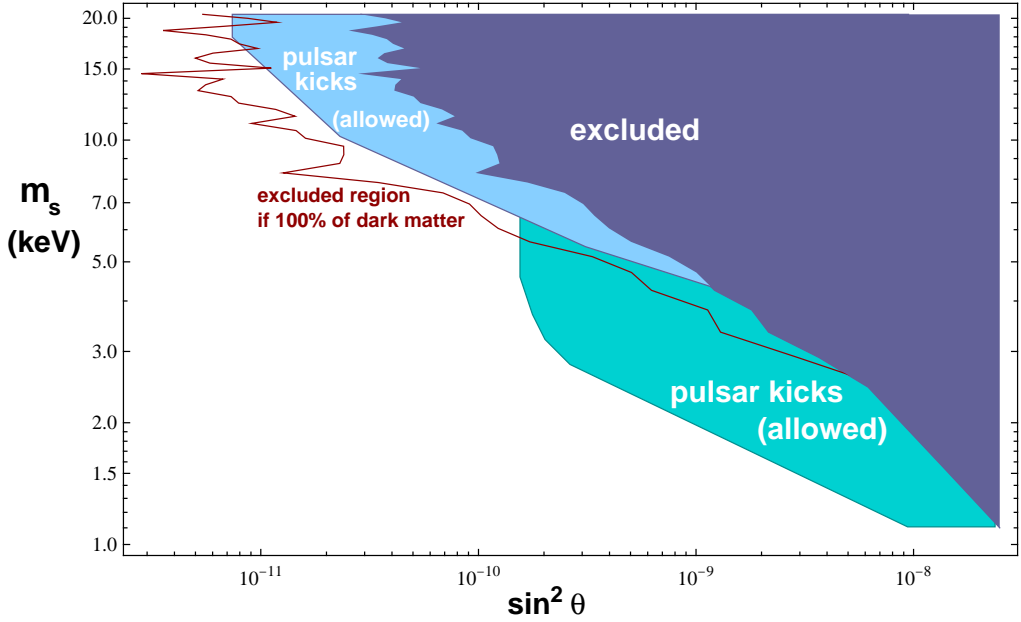}
  \caption{The solid excluded region is based on a combination of the X-ray 
and cosmological bounds; it applies even if sterile neutrinos constitute only a
fraction of dark matter~\cite{Kusenko:2006rh,Palazzo:2007gz,Loewenstein:2008yi}.
The dashed line shows the X-ray bound under the assumption that sterile
neutrinos (produced by some mechanism that may be different from oscillations)
make up all the dark matter.   Additional bounds from small-scale structure
formation (see discussion of Lyman-$\alpha$ observations) may apply, depending
on the free-streaming length, whose relation with the particle mass depends on
the
production scenario~\cite{Boyanovsky:2008nc,Petraki:2008ef,Gorbunov:2008ka,Boyarsky:2008xj,Boyarsky:2008mt,Wu:2009yr}. The allowed regions for the pulsar
kicks shown here  are based on
Refs.~\cite{Kusenko:1997sp,Kusenko:2008gh,Fuller:2003gy}.}
  \label{fig:range}
\end{figure}

 If all the dark matter is made up of sterile neutrinos
($\Omega_s\approx 0.2 $), then the limit on the mass and the mixing angle is
given by the dashed line in Fig.~\ref{fig:range}. 
However, the interactions in the Lagrangian (\ref{lagrangianM}) cannot produce
such an $ \Omega_s= 0.2 $ population of sterile neutrinos for the masses
and mixing angles along this dashed line, unless the universe has a relatively
large lepton asymmetry~\cite{Shi:1998km}.  If the lepton asymmetry is small,
the interactions included in the Lagrangian of eq.~(\ref{lagrangianM}) can produce the relic sterile
neutrinos only via the neutrino oscillations off-resonance at some sub-GeV
temperature~\cite{Dodelson:1993je}.  The model-independent
bound~\cite{Kusenko:2006rh,Palazzo:2007gz,Boyarsky:2008xj,Boyarsky:2008mt} based
on this scenario is shown as a solid (purple) region in Fig.~\ref{fig:range}. 
It is based on the
flux limit from X-ray observations~\cite{Abazajian:2001vt,Abazajian:2005gj,Beacom:2005qv,Mapelli:2005hq,Boyarsky:2005us,Abazajian:2006yn,Boyarsky:2006fg,Boyarsky:2006jm,Watson:2006qb,Abazajian:2006jc,Boyarsky:2006ag,Boyarsky:2006hr,RiemerSorensen:2006fh,RiemerSorensen:2006pi,Boyarsky:2006kc,Boyarsky:2006jm,Yuksel:2007xh,Boyarsky:2007ay,Loewenstein:2008yi} and the state-of-the-art calculation of the sterile neutrino production by oscillations~\cite{Asaka:2006rw,Asaka:2006nq}. 

Finally, there are differences in the way the data are analyzed in search of the line from the sterile neutrino decay.  
For example, some limits are based on the requirement that the X-ray flux from dark matter not exceed the observed total 
flux from a given object.  Some other limits are derived using the background modeling based on known astrophysical sources.  
Finally, yet another strategy is to use an \textit{ad hoc} numerical fit to the data and to require that the goodness of fit not be 
affected by the addition of a line at a given energy.  While each of these strategies can be defended, it is obvious that the former approach produces the most conservative limits, while the latter produces the strongest, but less robust limits.  

Instrumental backgrounds and energy resolution often limit one's ability to detect a line.   While some existing telescopes, such as, e.g.,  \textit{Suzaku} have a relatively stable background and reasonable energy resolution, the advent of new instruments, such as Astro-H and Constellation-X,  with a much better energy resolution will allow a much more efficient  search.   Meanwhile, the ongoing search using \textit{Suzaku} X-ray telescope targets the dark matter dominated dwarf spheroid galaxies, such as Draco and Ursa Minor~\cite{2008cxo..prop.2676L,2008HEAD...10.2906L}.  The outcome of this search may allow one to explore a substantial part of the range of parameters in which the pulsar kicks and dark matter could be explained simultaneously.  

\section{Early decays, X-rays, and the formation of the first stars}

 The X-ray photons from sterile neutrino decays in the early universe could
have affected the star formation.  Although these X-rays alone are not
sufficient to reionize the universe, they can catalyze the production of molecular hydrogen and speed up the star
formation~\cite{Biermann:2006bu,Stasielak:2006br,Stasielak:2007ex,Stasielak:2007vs}, which, in turn, would cause the reionization. Molecular hydrogen is a very important
cooling agent, necessary for the collapse of primordial gas clouds that gave
birth to the first stars.  The fraction of molecular hydrogen must exceed a
certain minimal value for the star formation to begin~\cite{Tegmark:1996yt}. 
The reaction 
\begin{eqnarray}
 {\rm H}+{\rm H} &\rightarrow & {\rm H}_2 +\gamma 
\end{eqnarray}
 is very slow in comparison with the combination of reactions: 
\begin{eqnarray}
{\rm H}^{+}+{\rm H}  & \rightarrow & {\rm H}_2^++ \gamma , \\  
{\rm H}_2^{+}+{\rm H} & \rightarrow & {\rm H}_2+{\rm H}^+, 
\end{eqnarray} %
which become possible if some hydrogen is ionized.  Thus, the ionization
fraction determines the rate of molecular hydrogen production.  If dark
matter is made up of sterile neutrinos, their decays can produce a sufficient flux
of photons to cause a significant increase in the ionization fraction~\cite{Biermann:2006bu,Stasielak:2006br,Stasielak:2007ex,Stasielak:2007vs}.  This can have a dramatic effect on the production of molecular hydrogen and the subsequent star formation.

Decays of the relic sterile neutrinos during the dark ages could produce an
observable signature in the 21-cm background~\cite{Valdes:2007cu}. It can be
detected and studied with such instruments as the Low Frequency Array (LOFAR),
the 21 Centimeter Array (21CMA), the Mileura Wide-field Array (MWA) and the
Square Kilometer Array~(SKA). 

\section{If this is right...}

If, indeed, the relic sterile neutrinos are discovered,  the sharp spectral line from their two-body radiative decay will be a boon to observational cosmology, because one can hope to get the distance--redshift information from the X-ray observations of dark matter halos.   In addition, the X-ray  instruments with high energy resolution and good imaging capabilities could be used to map out the dark matter halos that contain no luminous matter.  

\section{Conclusions}

The fundamental physics responsible for the neutrino masses is likely to involve some  gauge-singlet fermions.  Their Majorana masses can range from a few eV to some values well above the electroweak scale.  If some of the Majorana masses are below the electroweak scale,  the corresponding new degrees of freedom can have direct observable consequences.   In particular, a sterile neutrino with mass of several keV is a viable dark matter candidate~\cite{Dodelson:1993je}, and  the same particle can explain the pulsar kicks~\cite{Fuller:2003gy,Kusenko:1997sp} and can play a role in other astrophysical phenomena.  

The search for an X-ray line from the radiative decay of relic sterile neutrinos provides the best opportunity to discover these particles if they exist.  Depending on the model parameters, relic sterile neutrinos can constitute all or part of cosmological dark matter.   Therefore, dark matter dominated systems, such as dwarf spheroid galaxies, which are also X-ray quiet,  are among the best targets for current and planned X-ray telescopes.   In addition, the diffuse extragalactic background and the Milky Way halo can be used to search for sterile dark matter.  The improved energy resolution of the next generation of X-ray telescopes, such as Astro~H  (know as NeXT in its earlier stages of development), will help explore the best motivated region of parameters, in which both dark matter and the pulsar kicks are explained by the same sterile neutrino.  Due to the small mixing angles, the laboratory experiments will  probably not be competitive with the astrophysical searches in the foreseeable future.  However, if sterile neutrinos are to be a relatively cold dark matter, the production of these particles should have taken place at or above the electroweak scale, and LHC may be able to discover the requisite gauge singlet in the Higgs sector~\cite{Kusenko:2006rh,Petraki:2007gq}. The observations and the  population studies of pulsars can provide some additional clues regarding the role of sterile neutrinos in generating the pulsar kicks.   Given the broad range of X-ray searches that have been conducted and are being planned, one can hope to explore the best-motivated range of masses and mixing angles in the near future. 

\section{Acknowledgments}

This work was supported in part  by DOE grant DE-FG03-91ER40662 and by the
NASA ATFP grant  NNX08AL48G.  The author appreciates the
hospitality of Aspen Center for Physics, where part of this work was done.

\bibliographystyle{h-elsevier3}
\bibliography{sterile}

\end{document}